\documentclass[
    prl, twocolumn, superscriptaddress, nofootinbib, amsmath, amssymb,
    aps, floatfix, preprintnumbers
]{revtex4-2}

\usepackage[utf8]{inputenc}
\usepackage{setspace}
\usepackage[dvips]{graphicx}
\usepackage[dvipsnames]{xcolor}
\definecolor{CiteBlue}{RGB}{45,52,151}
\usepackage[
    colorlinks=true,
    linkcolor=CiteBlue,
    urlcolor=CiteBlue,
    citecolor=CiteBlue
]{hyperref}
\usepackage{aas_macros}

\usepackage[capitalise]{cleveref}
\usepackage[version=4]{mhchem}
\usepackage{siunitx}
\DeclareSIUnit{\year}{yr}

\let\oldsection\section

\newcommand{\refcite}[1]{Ref.~\cite{#1}}
\newcommand{\refscite}[1]{Refs.~\cite{#1}}

\usepackage{bm}
\newcommand{\bb}[1]{\bm{\mathrm{#1}}}
\newcommand{\du}{\mathrm{d}}
\newcommand{\dd}{\,\du}
\renewcommand{\Im}{\operatorname{Im}}
\renewcommand{\Re}{\operatorname{Re}}
\newcommand{\dm}{\chi}
\newcommand{\med}{\phi}
\newcommand{\plas}{\mathrm{p}}
\newcommand{\lindhard}{\mathrm{L}}
\newcommand{\el}{\mathrm{e}}
\newcommand{\fit}{\mathrm{fit}}
\newcommand{\rrscan}{r$^2$SCAN}
\newcommand{\materialsproject}{\textit{Materials Project}}
\newcommand{\mpr}{\texttt{MP}}
\newcommand{\mpid}[1]{\href{https://next-gen.materialsproject.org/materials/mp-#1}{\texttt{mp-#1}}}

\begin{document}
\title{First High-Throughput Evaluation of Dark Matter Detector Materials}
\preprint{MIT-CTP/5862}

\author{Sin\'{e}ad M. Griffin}
\affiliation{Materials Sciences Division, Lawrence Berkeley National Laboratory, Berkeley, CA 94720, USA}\affiliation{Molecular Foundry, Lawrence Berkeley National Laboratory, Berkeley, CA 94720, USA}

\author{Yonit Hochberg}
\affiliation{Racah Institute of Physics, Hebrew University of Jerusalem, Jerusalem 91904, Israel}
\affiliation{Laboratory for Elementary Particle Physics, Cornell University, Ithaca, NY 14853, USA}
 
\author{Benjamin V. Lehmann}
\affiliation{Center for Theoretical Physics -- a Leinweber Institute, Massachusetts Institute of Technology, Cambridge, MA 02139, USA}

\author{Rotem Ovadia}
\affiliation{Racah Institute of Physics, Hebrew University of Jerusalem, Jerusalem 91904, Israel}
\affiliation{Laboratory for Elementary Particle Physics,
 Cornell University, Ithaca, NY 14853, USA}

\author{Kristin A. Persson}
\affiliation{Materials Sciences Division, Lawrence Berkeley National Laboratory, Berkeley, CA 94720, USA}
\affiliation{Department of Materials Science and Engineering, University of California, Berkeley, Berkeley, CA, 94720 USA}

\author{Bethany A. Suter}
\affiliation{Leinweber Institute for Theoretical Physics, University of California, Berkeley, CA 94720, USA}

\author{Ruo Xi Yang}
\affiliation{Materials Sciences Division, Lawrence Berkeley National Laboratory, Berkeley, CA 94720, USA}
\affiliation{Molecular Foundry, Lawrence Berkeley National Laboratory, Berkeley, CA 94720, USA}

\author{Wayne Zhao}
\affiliation{Materials Sciences Division, Lawrence Berkeley National Laboratory, Berkeley, CA 94720, USA}
\affiliation{Department of Materials Science and Engineering, University of California, Berkeley, Berkeley, CA, 94720 USA}
\affiliation{Liquid Sunlight Alliance and Chemical Sciences Division, Lawrence Berkeley National Laboratory, Berkeley, 94720, CA, USA}

\date\today

\begin{abstract}\ignorespaces{}
    We perform the first high-throughput search and evaluation of materials that can serve as excellent low-mass dark matter detectors. Using properties of close to one thousand materials from the \materialsproject{} database, we project the sensitivity in dark matter parameter space for experiments constructed from each material, including both absorption and scattering processes between dark matter and electrons. Using the anisotropic materials in the dataset, we further compute the level of daily modulation in interaction rates and the resulting directional sensitivities, highlighting materials with prospects to detect the dark matter wind. Our methods provide the basic tools for the data-driven design of dark matter detectors, and our findings lay the groundwork for the next generation of highly optimized direct searches for dark matter as light as the keV scale.
\end{abstract}

\maketitle

The identity of the dark matter (DM) in the Universe remains a pressing open question. Decades of experimental searches targeting candidates near the weak scale have yielded null results, which has heightened interest in scenarios with masses well away from the weak scale~\cite{Essig:2011nj}. Over the last ten years, a robust community effort has emerged to develop experiments capable of probing DM scattering for masses well below a GeV, even down to the keV scale, where cosmological constraints become severe. This program has motivated the study of a wide range of detection modalities, and, in particular, various detector constituents. Proposed detector materials range from the familiar to the exotic, including scintillators~\cite{Derenzo:2016fse}, semiconductors~\cite{Essig:2015cda,Graham:2012su,Hochberg:2016sqx,Griffin:2020lgd}, superconductors~\cite{Hochberg:2015pha,Hochberg:2019cyy,Hochberg:2021yud,Gao:2024irf}, superfluid~\cite{Schutz:2016tid,Knapen:2016cue} and solid helium~\cite{Ashour:2024xfp}, two-dimensional systems~\cite{Hochberg:2016ntt,Cavoto:2017otc}, aromatic organic compounds~\cite{Blanco:2021hlm}, and even solid ice~\cite{Taufertshofer:2023rgq}. Several of these proposals have already been realized in the laboratory, placing new limits on light DM~\cite{Hochberg:2019cyy,Hochberg:2021yud,SENSEI:2023zdf,Gao:2024irf,SuperCDMS:2024yiv}.

Given the vast array of detection strategies with widely varying properties, a key component of this program is the identification of the modalities and materials that will enable the most sensitive experiments. The structure of the detector system, including its constituent materials, determines the nature and properties of the excitations that can be produced by DM interactions. Much of the early work on sub-GeV direct detection focused on identifying systems with detectable low-energy excitations: the detection of light DM is ultimately limited by the energy budget of the incoming DM, which is $\mathcal O(\qty{}{\milli\electronvolt})$ for elastically-scattering keV-scale DM\@. Designing a system to read out such a small excitation is highly nontrivial. This was, for example, the original motivation for superconductors as DM detectors: since typical superconductors have a gap of $\mathcal O(\qty{}{\milli\electronvolt})$, they exhibit a set of low-energy excitations suitable for light DM detection.

With experimental technologies now maturing, material selection has come into the limelight. When DM interacts with a detector, the entire detector system can respond to the induced perturbation. The nature of this detector response plays a key role in determining the DM interaction rate at a given energy. Thus, to design an experiment sensitive to light DM, one should aim to optimize the entire energy- and momentum-dependent detector response in the corresponding regime, which is sensitive to all collective modes of the system. In general, this is a complicated task, since detector response can be difficult to compute. However, for the specific case of spin-independent DM-electron interactions, the role of detector response is now well understood~\cite{Hochberg:2021pkt,Knapen:2021run,Boyd:2022tcn}. A single `loss function' can be defined in terms of the energy and momentum transfer in the scattering process, and this loss function fully determines the DM interaction rate. Since the loss function in certain regimes also determines the scattering rate of Standard Model particles in the material, it can be measured experimentally or computed using standard tools from the condensed matter community. This opens an opportunity anticipated four years ago by \refcite{Hochberg:2021pkt}: that one could rapidly survey a large number of materials for suitability in direct detection experiments.

In this work, we realize this analysis for the first time. We use results from the \materialsproject{}\footnote{Publicly available at \href{https://next-gen.materialsproject.org}{\texttt{next-gen.materialsproject.org}}.}~(\mpr)~\cite{Jain:2013wst,Petousis:2017,Ruoxi:2022}, a large database of state-of-the-art computations of inorganic material properties, including dielectric response functions, which give rise to the loss function that controls DM interaction rates. This large catalog of computed loss functions allows us to discover candidate materials in a data-driven way, with no prior notion of their properties. (Smaller-scale searches for spin-orbit gapped semiconductors~\cite{Inzani:2020szg} and organic semiconductors~\cite{Geilhufe:2018gry} have been considered in the context of low-mass DM targets, but these searches focused on computing target properties only, and did not evaluate their DM responses.) Moreover, since the \mpr{} dataset includes all components of the static dielectric tensor, we can also evaluate the suitability of anisotropic materials for directionally sensitive searches, a crucial step for background rejection in the next generation of experiments~\cite{Hochberg:2016ntt,Cavoto:2017otc,Hochberg:2017wce,Budnik:2017sbu,Kadribasic:2017obi,Griffin:2018bjn,Heikinheimo:2019lwg,Coskuner:2019odd,Geilhufe:2019ndy,Coskuner:2021qxo,Sassi:2021umf,Blanco:2021hlm,Hochberg:2021ymx,Blanco:2022pkt,Boyd:2022tcn}. Our work is the first large-scale study that promises to rapidly and dramatically expand the pool of viable DM detector materials by harnessing enormous computational and experimental datasets from beyond the confines of the DM community.

In this work, we use natural units with $c = \hbar = 1$.

\section{DM interaction rate}

To calculate the interaction rate between the DM and the electrons in the detector, we follow \refscite{Hochberg:2021pkt,Boyd:2022tcn}, which we now summarize briefly. We assume that the DM species $\dm$ has a spin-independent interaction with electrons in the detector. We denote the energy transferred in the interaction by $\omega$, and we denote the 3-momentum transferred by $\bb q$, with $q = \left|\bb q\right|$. The interaction rate is determined by Fermi's Golden Rule via the dynamic structure factor,
\begin{equation}
    \label{eq:dynamic-structure-factor-definition}
    S(\bb q, \omega) =
    \frac{2\pi}{V} \sum_f \bigl|
    \left\langle f\right|
        \hat n_{\el} \left( -\bb q \right) \left|0\right\rangle
    \bigr|^2 2 \pi \delta \left(E_f-E_i - \omega \right)
    ,
\end{equation}
where $V$ is the volume of the system, $\hat n_{\el}$ is the electron number density operator, and the sum is taken over all accessible detector final states $\left|f\right\rangle$. As discussed at length in \refscite{Hochberg:2021pkt,Boyd:2022tcn}, the dynamic structure factor can be rewritten in terms of the dielectric response function. In particular, given a dielectric tensor $\tensor\epsilon$, we have
\begin{equation}
    \label{eq:loss-function-tensor}
    S(\bb q, \omega) = \frac{8 \pi \alpha}{q^2}
        \Im\left(
            -\frac{1}{\bb{\hat q}
            \cdot\tensor{\epsilon}(\bb q, \omega)
            \cdot\bb{\hat q}}
        \right)
    ,
\end{equation}
where $\alpha\approx1/137$ is the fine structure constant. The quantity $\mathcal W(\bb q,\omega) \equiv \Im[-1/(\bb{\hat q}\cdot\tensor{\epsilon}(\bb q, \omega)\cdot\bb{\hat q})]$ is known as the loss function. The underlying premise of this work is that dielectric tensors have been computed for many different materials by the \mpr{}, and these can be used almost directly to predict the sensitivity of an experiment.

The rate of scattering between DM and an electronic target in the laboratory frame is given in terms of the dynamic structure factor by
\begin{equation}\label{eq:rate}
    R(t) = \frac{1}{\rho_{\mathrm{T}}} \frac{\rho_\dm}{m_\dm}
    \frac{\pi \bar\sigma_\el}{\mu_{\el\dm}^2}
    \int \frac{\du^3\bb q\dd\omega}{(2\pi)^3}
        g_0(\bb q, \omega, t) \mathcal F(q)^2 S(\bb q, \omega)
    .
\end{equation}
Here $\rho_\dm$ and $\rho_{\mathrm{T}}$ are the densities of the DM and the target, respectively; $\mu_{\el\dm}$ is the DM-electron reduced mass; $\mathcal F(q)$ is a form factor that characterizes the dependence on the mediator; $\bar\sigma_\el$ is a reference cross section; and $g_0(\bb q, \omega, t)$ characterizes the DM distribution in the kinematically accessible parameter space. Explicitly, 
\begin{equation}
    \label{eq:sigma-e}
    \mathcal F(q) =
        \frac{(\alpha m_\el)^2 + m_\med^2}{q^2 + m_\med^2}
    ,
    \quad
    \bar\sigma_\el =
        \frac{1}{\pi} \frac{\mu_{\el\dm}^2 g_{\el}^2 g_\dm^2}
            {\bigl[(\alpha m_\el)^2 + m_\med^2\bigr]^2}
    ,
\end{equation}
with $m_\med$ the mediator mass, and $g_0$ is given by
\begin{equation}
    \label{eq:dm-distribution}
    g_0(\bb q, \omega, t) =
    \int\du^3\bb v_\dm \, f_\dm^{\mathrm{lab}}(\bb v_\dm, t)
        \, \delta\left[\omega - \omega_{\bb q}(\bb v_\dm)\right]
    ,
\end{equation}
where $\omega_{\bb q}(\bb v_\dm) = \bb{v_\dm} \cdot \bb q - q^2/(2 m_\dm)$ and $f_\dm^{\mathrm{lab}}(\bb v_\dm, t)$ is the 3D DM velocity distribution in the laboratory frame. Note that $f_\dm$ is time-dependent due to the rotation of the Earth with respect to its direction of motion through the Galactic halo. The changing direction of the DM `wind' in the laboratory frame translates to a daily modulation in the rate for detectors with anisotropic dielectric tensors~\cite{Spergel:1987kx,Boyd:2022tcn}. This directional sensitivity allows for the detection of the DM wind, enabling strong background rejection and confirmation of a putative discovery~\cite{Mayet:2016zxu}. We use the standard halo model~\cite{Lewin:1995rx} for $f_\dm^{\mathrm{lab}}$, with dispersion velocity $v_0 = \qty{220}{\kilo\meter/\second}$, escape velocity $v_\mathrm{esc} = \qty{550}{\kilo\meter/\second}$, and Earth velocity $v_{\mathrm{E}} = \qty{232}{\kilo\meter/\second}$.

\begin{figure*}\centering
    \includegraphics[width=\textwidth]{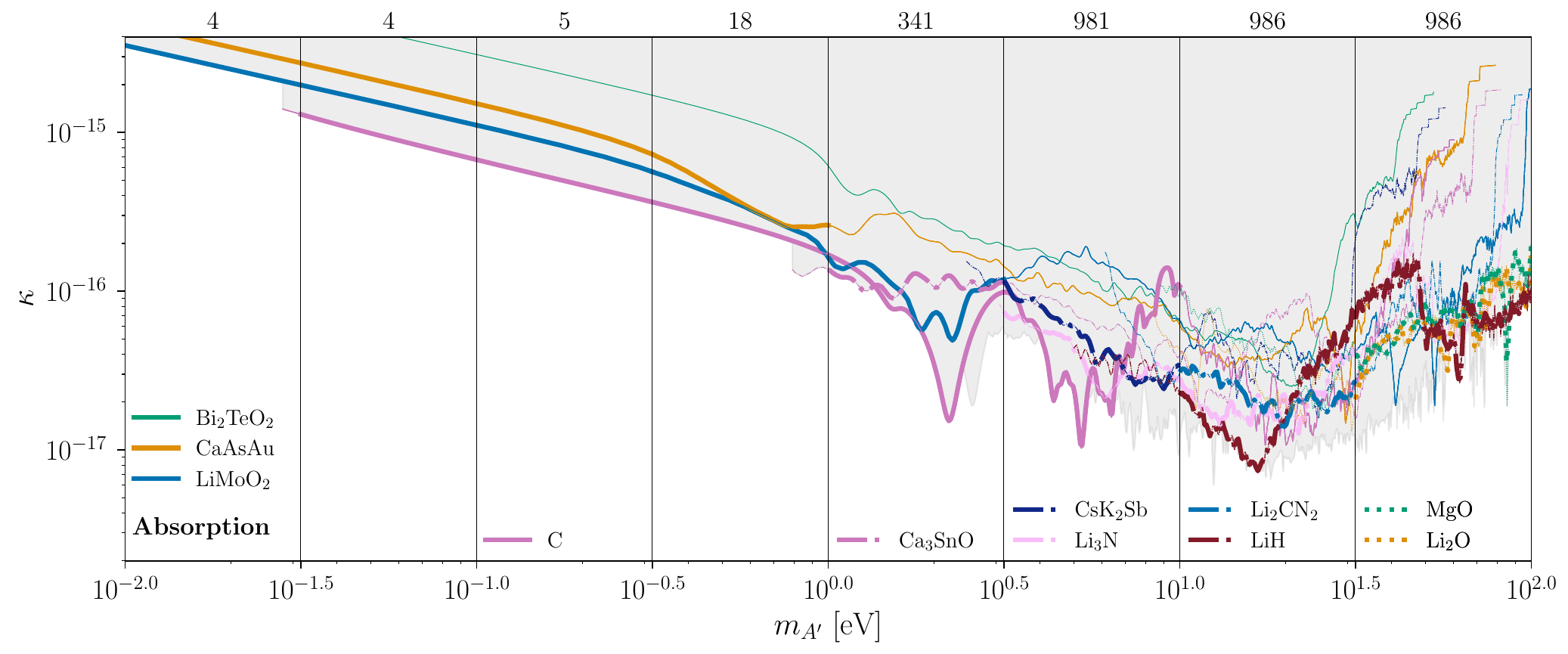}
    \caption{\textbf{Absorption.} Projected reach of optimal materials from the \mpr{} dataset for DM absorption of a kinetically mixed dark photon. In each half-decade of DM mass, materials are ranked by average reach in log space, and the three best  from the dataset are indicated by thick lines in each panel. The gray region indicates parameter space that is accessible by the combination of all materials in the \mpr{} dataset. The numbers across the top indicate the number of materials with any sensitivity in each mass range. All projections are made at 95\% CL for a background-free exposure of 1 kg-yr.}
    \label{fig:absorption}
\end{figure*}

We also consider DM absorption for the case of a kinetically mixed dark photon, \textit{i.e.}, a model with an interaction of the form $\mathcal L \supset -\frac12\kappa F_{\mu\nu}F'^{\mu\nu}$, where $F$ and $F'$ are the field strength tensors for the Standard Model photon and dark photon, respectively. This interaction will occur at a rate mimicking Standard Model photon absorption in a material, but suppressed by a factor of $\kappa^2$. Thus, the absorption rate per unit detector volume per DM particle is given by~\cite{Knapen:2021bwg, Hochberg:2016sqx}
\begin{equation}
    \label{eq:isotropic-absorption}
    \Gamma_{\mathrm{A}} = \kappa^2m_\dm
        \Im\left(-\frac{1}{\epsilon(m_\dm \bb v_\dm,\,m_\dm)}\right)
    .
\end{equation}
Since $|\bb v_\dm|\sim\num{e-3}$, \cref{eq:isotropic-absorption} only samples the loss function in the regime $|\bb q| \ll \omega$. Thus, when computing the absorption rate, we fix $|\bb q| = 0$, which gives an excellent approximation for the DM masses considered in this work. Here we assume that the dielectric tensor is isotropic, and write $\epsilon$ in place of $\tensor\epsilon$. When $\tensor\epsilon$ is anisotropic, we average over Cartesian directions. The computation of the DM absorption rate in an anisotropic material will be the subject of future work~\cite{to-appear}.

\begin{figure*}\centering
    \includegraphics[width=\textwidth]{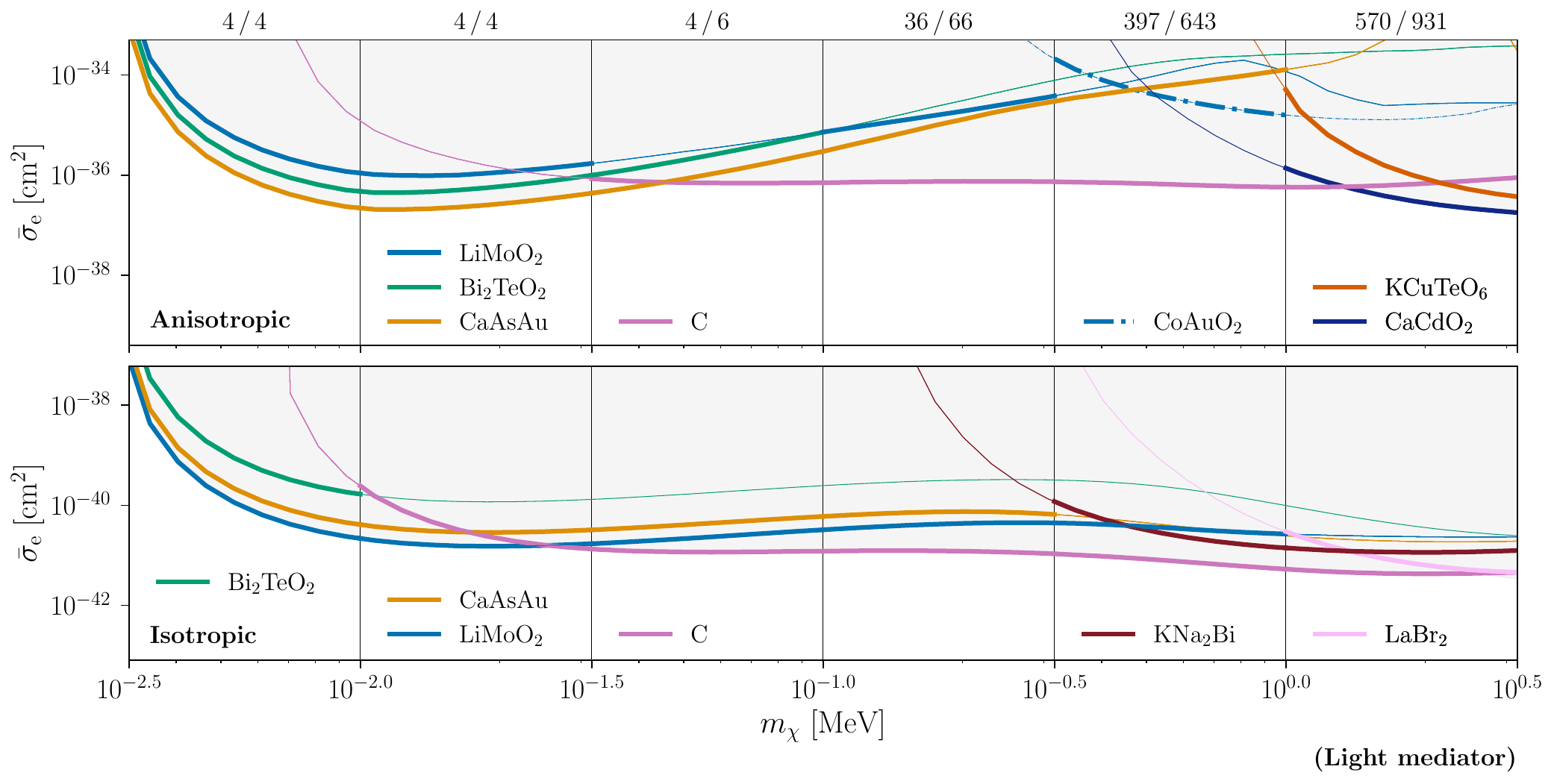}
    \caption{\textbf{Scattering via a light mediator.} Projected reach of optimal materials in the \mpr{} dataset for DM scattering via a light mediator, selected on the same criteria as in \cref{fig:absorption}.\ \textit{Bottom:} optimal materials in the \mpr{} dataset on the basis of 3-event reach.\ \textit{Top:} optimal materials on the basis of sensitivity to anisotropy in the DM distribution. The pairs of numbers across the top indicate the number of materials with any sensitivity in each mass range: the first and second numbers correspond to anisotropic and isotropic sensitivity, respectively. All projections are made at 95\% CL for a kg-yr exposure.}
    \label{fig:scattering-light}
\end{figure*}

\section{Dataset and fitting procedure}

The \mpr{} dataset~\cite{Jain:2013wst,Petousis:2017,Ruoxi:2022} includes dielectric tensors, band gaps, densities, and unit cell volumes. We use improved band gap computations, some of which are not yet incorporated into the \mpr{} dataset, and we consider \num{986} out of the original \num{1019} materials for which reliable band gaps are available. Further details are given in the Supplemental Material (SM), and values are given in a public repository released together with this work~\cite{full_repository}.

In the \mpr{} dataset, the dielectric tensor $\tensor\epsilon_{\mathrm{data}}$ is only available in the limit $\bb q\to0$. This is sufficient to treat DM absorption, which always takes place in the small-$q$ limit, but it is insufficient for DM scattering. It is therefore necessary to adopt a fixed procedure for estimating the dielectric tensor at $q>0$. We use the following steps. First, if $\tensor\epsilon_{\mathrm{data}}$ is anisotropic, we average $\bb{\hat q}\cdot\tensor\epsilon_{\mathrm{data}}\cdot\bb{\hat q}$ over Cartesian directions to obtain a scalar dielectric function $\epsilon_{\mathrm{data}}$ for the purposes of estimating the overall rate.\footnote{We use the full anisotropic dielectric tensor for the purposes of estimating rate modulations, as described in the next subsection. The degree of anisotropy has minimal impact on the average rate.} Next, we extract all energies $\hat\omega_k$ where $\Re(\epsilon)$ vanishes. Each of these zeros except the first, $\hat\omega_0$, corresponds to a plasmon peak in the loss function. We fit each of these peaks with a Lindhard dielectric function $\epsilon_\lindhard$~\cite{Dressel_Gruner:2002}, fixing the plasma frequency to $\omega_{\plas,k} = \hat\omega_k$ and fitting the plasmon width $\Gamma_{\plas,k}$. We then construct a $q$-dependent loss function as the linear combination of these fits:
\begin{equation}
    \label{eq:combined-fit}
    \mathcal{W}_{\fit}(\bb q, \omega) =
    \frac{1}{\sum_{k} h_k}\sum_{k=1}^{n_{\mathrm{peaks}}}
        h_k \, \Im \left( - \frac{1}{\epsilon_\lindhard(\omega_{\plas, k},
            \Gamma_{\plas,k};\;\bb q, \omega)}\right) \, .
\end{equation}
(Sums of model loss functions have been previously used to approximate losses in real materials \textit{e.g.} by~\refscite{Abril:1998rea,2019JPCS..124..242V}.) This does not reduce to the loss function constructed from $\epsilon_{\mathrm{data}}$ as $q \to 0$. Thus, when defining the $q$-dependent loss function, we incorporate the residual ratio between the data and the fit to enforce agreement at $q = 0$, and we make the approximation that this ratio is independent of $q$. Specifically, we define $r(\omega)\equiv \Im[-1/\epsilon_{\mathrm{data}}(\omega)] / \left.\mathcal{W}_\fit(\bb q, \omega) \right|_{\bb q \rightarrow 0}$, and use a modified loss function
\begin{equation}
    \mathcal W_{r}(\bb q, \omega) =
    r(\omega)
    \times
    \mathcal{W}_{\fit}(\bb q, \omega)\,,
    \label{eq:fit-with-ratio}
\end{equation}
which retains the momentum dependence implied by the fitted Lindhard functions, but is constructed to yield $\mathcal W_{\mathrm{data}}$ in the limit $q \to 0$. We use $\mathcal W_{r}$ as a proxy for the material's loss function at $q > 0$ in our rate computations. While this empirical fitting procedure does not arise from a microphysical model, it is nonetheless capable of reproducing results from dedicated density functional theory~(DFT) computations in other materials~\cite{Dino:future}. We exclude 55 materials due to incompatibility with our fitting procedure, all with band gaps above \qty{1}{\electronvolt}. We thus evaluate a total of 931 materials for isotropic scattering sensitivity.

To calculate the daily modulation of the DM scattering rate induced by the rotation of the Earth, we use a directionally-dependent generalization of the fitting procedure described above: we separately fit $\bb{\hat q}\cdot\tensor\epsilon_{\mathrm{data}}\cdot\bb{\hat q}$ for \num{800} uniformly sampled directions $\bb{\hat q}_i$, with the plasmon parameters and ratio factors $r(\omega)$ computed independently from the longitudinal dielectric tensor along each direction. We perform this directionally-dependent fit for materials in which the loss function differs by at least 25\% on average between two Cartesian axes. Of the \num{931} materials evaluated for scattering sensitivity, \num{593} exhibit this level of anisotropy. To avoid sensitivity to artificial anisotropy introduced by the fitting procedure, we check for the presence of approximate uniaxial symmetry in any of the material axes, and find such a symmetry in 527 of the anisotropic materials. For these materials, we enforce the rotational symmetry by averaging the components of $\tensor\epsilon_{\mathrm{data}}$ around the symmetry axis prior to fitting. Of the \num{593} anisotropic materials, \num{23} were incompatible with our fitting procedure and were excluded, all with band gaps above \qty{1}{\electronvolt}. We compute the modulation following \refcite{Boyd:2022tcn}, and when evaluating $\mathcal{W}_{\fit}(\bb q)$, we use the fit in the direction $\bb{\hat q}_i$ closest to $\bb{\hat q}$. Uniaxial materials are oriented with their axis perpendicular to the DM wind. For other materials, one of the Cartesian axes is oriented in this way, chosen to maximize the modulation. Further details of the selection criteria and preprocessing steps are given in the SM\@.

\begin{table*}
    \centering
    \resizebox{\textwidth}{!}{%
    \begin{tabular}{
        |llll|
        llll|
    }
        \hline\hline
        \textbf{Formula} & \textbf{MPID} & $\bm{E_g}\,[\qty{}{\electronvolt}]$ & \textbf{Structure type}&
        \textbf{Formula} & \textbf{MPID} & $\bm{E_g}\,[\qty{}{\electronvolt}]$ &
        \textbf{Structure type}
        \\
        \hline
        \ce{Bi2TeO2} & \mpid{755419} & Gapless
            & Anti-\ce{ThCr2Si2}~\cite{2015JSSCh.226..219L}
        & \ce{KNa2Bi} & \mpid{863707} & 0.43\,(H)
            & \ce{DO3}-type~\cite{1963AcCry..16..316S} 
        \\
        \ce{C} & \mpid{569304} & 0.21\,(r)
            & AABBCC graphite~\cite{1966RSPSA.291..324N} &
        \ce{LaBr2} & \mpid{28572} & 1.00\,(H)
            & 2\ce{H2}-\ce{MoS2}~\cite{Kramer:1989} 
        \\
        \ce{Ca3SnO} & \mpid{29241} & 0.40\,(H)
            & Inverse perovskite~\cite{Widera:1980} & 
        \ce{Li2CN2} & \mpid{9610} & 4.93\,(H)
            & Tetragonal I4/mmm~\cite{Down:1978} 
        \\
        \ce{CaAsAu} & \mpid{3927} & Gapless
            & Hexagonal $\omega$~\cite{Johrendt:1996} & 
        \ce{Li2O} & \mpid{1960} & 6.66\,(H)
            & \ce{CaF2} (fluorite)~\cite{Bijvoet:1926} 
        \\
        \ce{CaCdO2} & \mpid{753287} & 1.12\,(r)
            & Caswellsilverite-like &
        \ce{Li3N} & \mpid{2251} & 2.03\,(H)
            & Layered P6/mmm~\cite{Rabenau:1976} 
        \\
        \ce{CoAuO2} & \mpid{997161} & 0.63\,(r)
            & \ce{CuFeO2} (delafossite) &
        \ce{LiH} & \mpid{23703} & 3.99\,(H)
            & NaCl (rocksalt)~\cite{1972PhRvB...5.4704Z}
        \\
        \ce{Cs2O} & \mpid{7988} & 1.34\,(r)
            & Anti-\ce{CdCl2}/Layer~\cite{Tsai:1956} &
        \ce{LiMoO2} & \mpid{19338} & Gapless
            & Layered caswellsilverite~\cite{Aleandri:1988} 
        \\
        \ce{CsK2Sb} & \mpid{581024} & 1.67\,(H)
            & \ce{M3Sb} (M:alkali metal)~\cite{McCarroll:1965} & 
        \ce{Mg(BeN)2} & \mpid{11917} & 5.4\,(H)
            & \ce{La2O3}~\cite{Somer:2004}
        \\
        \ce{K2PdBr6} & \mpid{1205761} & 1.29\,(H)
            & \ce{K2PtCl6}~\cite{Ketblaar:1938} & 
        \ce{MgO} & \mpid{1265} & 6.62\,(H)
            & NaCl (rocksalt)~\cite{Broch:1927}
        \\
        \ce{KCuTeO6} & \mpid{1147551} & 2.0\,(H)
            & Layered
        & \ce{SrAlSiH} & \mpid{570485} & 0.63\,(G)
            & Layered Zintl hydride~\cite{2008PhRvB..78s5209L}
        \\
        \hline\hline
    \end{tabular}}
    \caption{Materials and their corresponding MPIDs studied in this work, listed alphabetically by column. For each material, we report the computed band gap $E_g$ and method: H (HSE hybrid functional), r (\rrscan\ meta-GGA), or G (GGA). We give the structure type for each material, and provide a reference when the compound listed has been experimentally synthesized. Materials listed as `Gapless' exhibited vanishing band gaps in each of the available computational methods.}
    \label{tab:materials}
\end{table*}
\section{Results and discussion}

We can now evaluate the sensitivity of a fiducial experiment to DM scattering and absorption for each of the materials in the \mpr{} dataset. For materials with anisotropic dielectric tensors, we further evaluate the amplitude of the daily modulation induced in the scattering rate. We have made these projections available for all of the materials in our dataset~\cite{full_repository}, with further details provided in the SM\@.

In order to identify the materials in the dataset with optimal sensitivity for DM searches, one needs a figure of merit for comparison. Given an interval of DM masses $(m_1, m_2)$, and a quantity $\alpha$ characterizing the strength of the weakest detectable interaction for a fixed exposure, we compute the figure of merit $\eta \equiv -\int_{\log m_1}^{\log m_2}\log\alpha(m_\dm)\dd\log m_\dm$, so that the largest value of $\eta$ corresponds to the system which is sensitive to the largest parameter space volume in logarithmic coordinates. For scattering, we take $\alpha$ to be the cross section $\bar\sigma_{\el}$ defined in \cref{eq:sigma-e}, and for absorption, we take $\alpha$ to be the kinetic mixing parameter $\kappa$. Ranking materials by the figure of merit $\eta$, we can identify the materials with greatest sensitivity in each mass range.\footnote{Note that only a small portion of the full \mpr{} database includes dielectric responses so far, so we describe materials as `optimal' only insofar as they are the best performers in the current dataset. Indeed, other materials not included in our dataset are known to exhibit better isotropic sensitivity in some mass ranges. Future application of our pipeline to improved versions of the dataset may reveal materials with better sensitivity as the dataset evolves.}

Our results for DM absorption of a kinetically mixed dark photon are shown in \cref{fig:absorption}. In each half-decade of DM mass, we show the projected sensitivity for all materials that are among the three materials with the largest figure of merit in the corresponding bin. The gray shaded region shows the best projected reach across all materials in the \mpr{} dataset at each DM mass. All projections are made at the 95\% confidence level (CL) assuming a background-free exposure of 1 kg-yr. We consider deposits $\omega > 10 \, \mathrm{meV}$, with the maximal deposit set by the highest energy available in the data for each material, at most \qty{100}{\electronvolt}. We label curves with their chemical formulae for readability, but these formulae do not uniquely determine material properties, since each formula can correspond to many different allotropes. An unambiguous description of a material is provided by its ID in the \mpr{} database, or MPID\@. \Cref{tab:materials} lists all materials appearing in our plots, including their MPID, relevant band gap, structure type, and whether they have been experimentally synthesized. 

Our results for DM scattering are shown in \cref{fig:scattering-light} for the case of a light mediator. The analogous results for a heavy mediator are given in the SM\@. These are qualitatively similar and select nearly the same set of optimal materials. As with \cref{fig:absorption}, the lower panel shows the three optimal materials from the \mpr{} dataset in each half-decade of DM mass. The top panel shows optimal materials for the directional detection of the DM wind. For a fixed exposure, this `anisotropic reach' is defined by the cross section for which the null hypothesis of an isotropic DM distribution can be rejected at 95\% CL\@. The anisotropic reach always corresponds to larger cross sections, since many events are required to establish that the arrival directions are not uniformly distributed. The anisotropic reach is representative of the best possible sensitivity in the presence of uncontrolled backgrounds, where the detection of the DM wind direction is crucial in order to distinguish the DM signal from other sources.

A key difference between our results for absorption and scattering is the range of masses over which particular materials are optimal. The kinematics of DM absorption involves an energy transfer on the order of the DM mass, which is much larger than the momentum transfer. Accordingly, localized features of the response function result in localized features in the reach curve, so a material that is optimal in one half-decade of mass is not generically optimal elsewhere. On the other hand, the kinematics of scattering average the response over a range of energy transfers, leading to smoother reach curves, and to materials which are optimal over wide mass ranges. 

\begin{figure}\centering
    \includegraphics[width=\columnwidth]{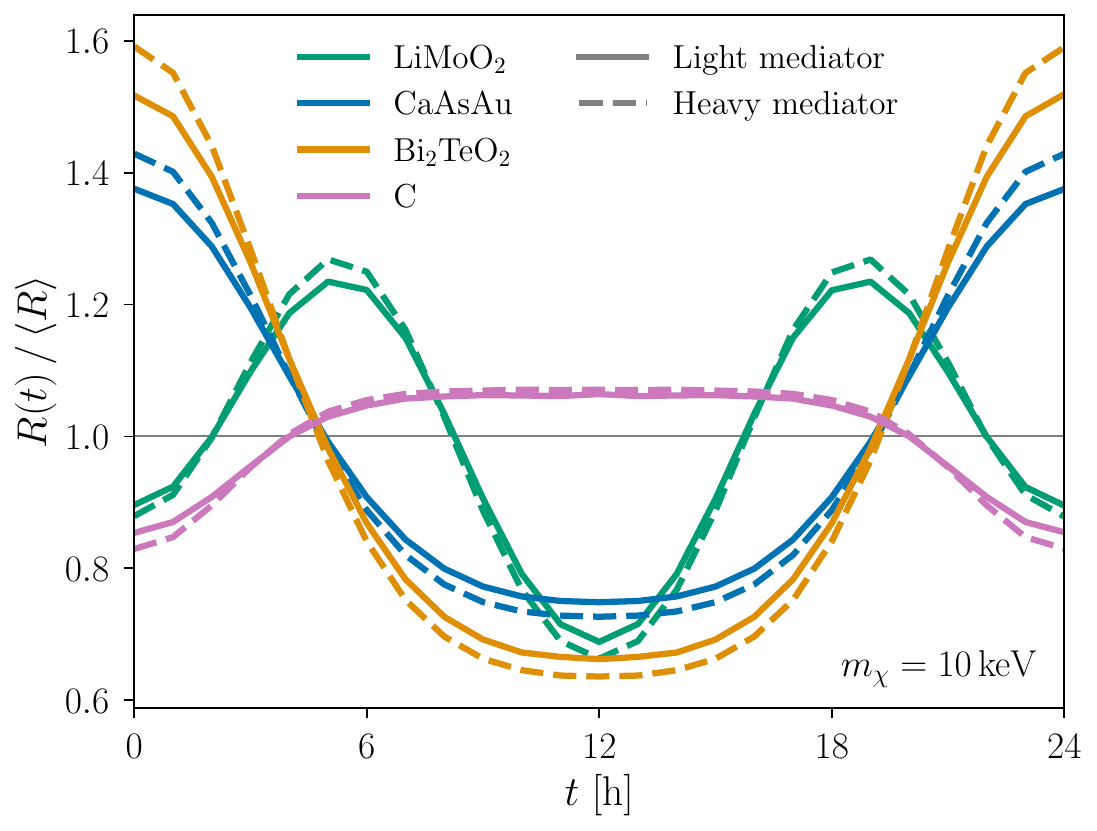}
    \caption{Daily modulation of the DM scattering rate for four selected materials calculated for $m_\dm = \qty{10}{\kilo\electronvolt}$. Solid (dashed) lines correspond to a light (heavy) mediator. The materials \ce{CaCdO2}, \ce{CoAuO2}, and \ce{KCuTeO6} exhibit daily modulations of similar order at $m_\chi \sim \qty{}{\mega\electronvolt}$.}
    \label{fig:daily-modulation}
\end{figure}

One might hope to optimize DM searches by selecting materials that offer excellent reach for both absorption and scattering simultaneously, and for the widest possible mass range. A small number of materials are among the optimal three across multiple half-decades of DM mass. For absorption, these materials are \ce{C}, \ce{LiMoO2}, and \ce{Li3N}. For isotropic scattering, these materials are \ce{C}, \ce{LiMoO2}, \ce{CaAsAu}, and \ce{KNa2Bi}. For anisotropic scattering, these are \ce{C}, \ce{LiMoO2}, \ce{CaAsAu}, \ce{Bi2TeO2}, and \ce{CaCdO2}. We show the modulation in the DM scattering rate over 24 hours for some of these materials in \cref{fig:daily-modulation}.

Three materials are of particular interest: \ce{LiMoO2}, \ce{CaAsAu}, and \ce{Bi2TeO2} offer strong performance for each of absorption, isotropic scattering, and anisotropic scattering at the lowest DM masses, with daily modulations of $\mathcal O(20\textnormal{--}50\%)$.\ \ce{LiMoO2} (lithium molybdenate) is well known~\cite{Barker:2003a,Barker:2003b,ben-kamel:2012} and has been shown to intercalate \ce{Li}.\ \ce{Bi2TeO2}, a bismuth oxychalcogenide, is known as a thermoelectric material. It has been studied theoretically by \refcite{2024JPEn....6b5011F}, and is commercially available.\ \ce{CaAsAu}, while not widely studied, has been synthesized~\cite{Johrendt:1996}, and \refcite{2019Natur566475Z} identifies it as a topological semimetal. (Note that the flag \texttt{Predicted Stable} on the \mpr{} is one of several descriptors of synthesizability and does not necessarily indicate that a material can or cannot be synthesized.) Our results motivate experimental evaluation of these materials as targets for DM searches. Detailed DFT computations of their loss functions at nonvanishing momenta will also be pursued in future work~\cite{futureDFT}.

While several detector materials have thus far been evaluated or calibrated with dielectric response data, the materials that we have identified as optimal within our dataset are unique in that they are the first to have been selected by data-driven techniques. Our work constitutes the first high-throughput search and evaluation of DM responses in materials datasets for DM detector materials, and represents the natural next step in the use of material data for DM experiments. Now that large volumes of data are available, the data-driven approach can be used not only for detector calibration, but also for material discovery. (See also \refcite{Cook:2024cgm} for recent progress in machine-learning-driven methods for material selection.) As the \materialsproject{} and other large-scale efforts continue to expand their datasets, the DM direct detection community will have the immediate opportunity to identify new classes of materials to accelerate DM detection well below the GeV scale.

\bigskip

\begin{acknowledgments}
\textbf{Acknowledgments.} We thank Roni Ilan, Zohar Ringel and Bashi Mandava for useful discussions, and Dino Novko and Antonio Politano for valuable input. We thank Aaron D. Kaplan for contributions to the computation of HSE band gaps used in this work. The color palette for figures in this work was selected programmatically\footnote{During this work, we developed a package, \texttt{MonteColor}~\cite{montecolor}, for creating custom colorblind-safe palettes.\ \texttt{MonteColor} is publicly available at \href{https://github.com/benvlehmann/montecolor}{\texttt{github.com/benvlehmann/montecolor}}.} to minimize overlap for several types of color vision deficiency using the color model of \refcite{Luo:2006}. We thank Leor and Maya Kuflik for valuable comments on color selection. S.M.G., K.A.P. and R.Y. acknowledge support from the U.S. Department of Energy, Office of Science, Office of Basic Energy Sciences, Materials Sciences and Engineering Division under contract No. DE-AC02-05-CH11231 (Materials Project program KC23MP). Work at the Molecular Foundry was supported by the Office of Science, Office of Basic Energy Sciences, of the U.S. Department of Energy under Contract No. DEAC02-05CH11231. The work of Y.H. is supported in part by the Israel Science Foundation (grant No.\ 1818/22) and by the Binational Science Foundation (grants No.\ 2018140 and No.\ 2022287). The work of Y.H. and R.O. is supported by an ERC STG grant (``Light-Dark,'' grant No.\ 101040019). The work of R.O.\ is further supported by the Milner Fellowship, the Binational Science Foundation Travel Grant No.\ 308300002, and the gracious hospitality of Cornell University. The work of B.V.L.\ is supported by the MIT Pappalardo Fellowship. The work of B.A.S.\ is supported by the NSF GRFP Fellowship and BSF-2018140. W.Z. acknowledges support from the Liquid Sunlight Alliance, which is supported by the U.S. Department of Energy, Office of Science, Office of Basic Energy Sciences, Fuels from Sunlight Hub under the award number DE-SC0021266. This project has received funding from the European Research Council (ERC) under the European Union’s Horizon Europe research and innovation programme (grant agreement No.\ 101040019).  Views and opinions expressed are however those of the author(s) only and do not necessarily reflect those of the European Union. The European Union cannot be held responsible for them. This material is based upon work supported by the National Science Foundation Graduate Research Fellowship Program under Grant No.\ DGE 2146752. Any opinions, findings, and conclusions or recommendations expressed in this material are those of the author(s) and do not necessarily reflect the views of the National Science Foundation.
\end{acknowledgments}

\bibliography{references}

%%%%%%%%%% Supplemental materials %%%%%%%%%%

\onecolumngrid{}
\clearpage

\setcounter{page}{1}
\setcounter{equation}{0}
\setcounter{figure}{0}
\setcounter{table}{0}
\setcounter{section}{0}
\setcounter{subsection}{0}
\renewcommand{\theequation}{S.\arabic{equation}}
\renewcommand{\thefigure}{S\arabic{figure}}
\renewcommand{\thetable}{S\arabic{table}}
\renewcommand{\thesection}{\Roman{section}}
\renewcommand{\thesubsection}{\Alph{subsection}}
\newcommand{\ssection}[1]{
    \addtocounter{section}{1}
    \oldsection{\thesection.~~~#1}
    \addtocounter{section}{-1}
    \refstepcounter{section}
    \noindent\ignorespaces{}
}
\newcommand{\ssubsection}[1]{
    \addtocounter{subsection}{1}
    \subsection{\thesubsection.~~~#1}
    \addtocounter{subsection}{-1}
    \refstepcounter{subsection}
    \noindent\ignorespaces{}
}
\newcommand{\fakeaffil}[2]{$^{#1}$\textit{#2}\\}

\thispagestyle{empty}
\begin{center}
    \begin{spacing}{1.2}
        \textbf{\large 
            Supplemental Material:\\
            First High-Throughput Evaluation of Dark Matter Detector Materials
        }
    \end{spacing}
    \par\smallskip
    Sin\'{e}ad M. Griffin,\textsuperscript{1,\hspace{2pt}2}
    Yonit Hochberg,\textsuperscript{3,4}
    Benjamin V. Lehmann,\textsuperscript{5}
    Rotem Ovadia,\textsuperscript{3}\\
    Kristin A. Persson,\textsuperscript{1,6}
    Bethany A. Suter,\textsuperscript{7}
    Ruo Xi Yang\textsuperscript{1,\hspace{2pt}2}
    and Wayne Zhao\textsuperscript{1,\hspace{2pt}6,\hspace{2pt}8} 
    \par\smallskip
    {\small
        \fakeaffil{1}{Materials Sciences Division, Lawrence Berkeley National Laboratory, Berkeley, CA 94720, USA}
        \fakeaffil{2}{Molecular Foundry, Lawrence Berkeley National Laboratory, Berkeley, CA 94720, USA}
        \fakeaffil{3}{Racah Institute of Physics, Hebrew University of Jerusalem, Jerusalem 91904, Israel}
        \fakeaffil{4}{Laboratory for Elementary Particle Physics, Cornell University, Ithaca, NY 14853, USA}
        \fakeaffil{5}{Center for Theoretical Physics -- a Leinweber Institute,\\Massachusetts Institute of Technology, Cambridge, MA 02139, USA}
        \fakeaffil{6}{Department of Materials Science and Engineering,\\University of California, Berkeley, Berkeley, CA, 94720 USA}
        \fakeaffil{7}{Leinweber Institute for Theoretical Physics, University of California, Berkeley, CA 94720, USA}
        \fakeaffil{8}{Liquid Sunlight Alliance and Chemical Sciences Division,\\ Lawrence Berkeley National Laboratory, Berkeley, 94720, CA, USA}
        (Dated: \today)
    }

\end{center}
\par\smallskip

In this Supplemental Material, we give additional details of the dataset, our fitting procedure, and benchmarks for the fitted dielectric functions, and we provide additional results.

\ssection{Dataset}

Our dataset consists of the dielectric tensors $\tensor\epsilon$, band gaps, densities, and unit cell volumes of \num{1019} materials from the \materialsproject{} (\mpr{}). Of these, 15 lack associated density values and cannot be used to compute absorption or scattering rates, and an additional 18 are excluded due to unreliable band gap computations. We thus consider 986 materials in our DM absorption analysis. In the DM scattering analysis, 46 materials exhibit dielectric functions for which  $\Re(\epsilon)$ crossed zero fewer than two times. These dielectric functions cannot be modeled by our fitting procedure, described in detail in the following section, and are therefore excluded. An additional \num{9} materials in the isotropic scattering analysis and \num{7} materials in the anisotropic scattering analysis did not pass a goodness-of-fit test, described in the following section, and were thus removed. Additionally, \num{16} materials are incompatible with the directional fitting procedure and are thus excluded. This results in a final set of \num{931} materials evaluated for isotropic scattering and \num{570} for anisotropic scattering.

For each material, the full complex dielectric tensor $\tensor\epsilon_{\mathrm{data}}$ is provided in the dataset at $\textbf{q}=0$ as a $3 \times 3$ symmetric matrix for \num{2001} values of the energy transfer, $\omega$. We perform several preprocessing steps before using $\tensor\epsilon_{\mathrm{data}}$ in our analysis pipeline. This accounts for any differences between the data in our public repository~\cite{full_repository} and the original data in the \mpr{} repository~\cite{Jain:2013wst,Petousis:2017,Ruoxi:2022}. We now describe these steps in detail.

\begin{enumerate}
\item 
We apply an $\operatorname{SO}(3)$ rotation to approximately diagonalize the dielectric tensor of each material. These rotations were chosen to minimize the off-diagonal components of the rotated tensors. For most materials, this procedure effectively eliminates the off-diagonal elements, and for the remainder, it reduces them to less than 5\% of the diagonal components in magnitude. We then neglect all off-diagonal components. Since such a rotation corresponds to a reorientation of the crystal in the lab frame, this does not affect any physical observables, but it significantly reduces numerical artifacts in directional averaging. Furthermore, this simplifies the computation of the anisotropic reach by minimizing the number of directions required for interpolation and allowing easy identification of symmetry axes.

\item
Some materials in the dataset exhibit small but nonzero loss at $\omega = 0$, which is nonphysical, and which can have a disproportionate impact on the projected sensitivity to light DM\@. To eliminate these artifacts, we set all material responses at zero energy and momentum transfer to zero prior to fitting: $\mathcal{W}(\omega = 0, \textbf{q} = 0) = 0$.

\item 
Both the band gaps and the dielectric tensors in the \mpr{} repository are computed using density functional theory~(DFT) with the Perdew–Burke–Ernzerhof (PBE) generalized gradient approximation (GGA)~\cite{Perdew:1996pki}.  However, to more accurately capture the physical band gaps, we use two additional functionals, the \rrscan{} meta-GGA functional~\cite{2022PhRvM...6a3801K,Furness:2020} and the Heyd-Scuseria-Ernzerhof~(HSE) hybrid functional~\cite{Heyd:2003rlw,2006JChPh.125v4106K}, to identify an improved band gap $E_g^{\mathrm{new}}$. We compute HSE band gaps following the procedure described in \refcite{Zhao:2025}, which exhibits improved accuracy for band gaps in the energy range relevant for this work. Where available, we prioritize nonzero HSE band gaps over GGA and \rrscan{} band gaps. Where HSE band gaps are not available, we choose the larger of the GGA and \rrscan{} values. Materials that we label as gapless have vanishing gaps in all available computations. When the improved band gap $E_g^\mathrm{new}$ differs from the original GGA band gap $E_g^\mathrm{GGA}$, we shift the energies in the dielectric response data accordingly:
\begin{equation}
    \label{eq:gap-shift}
    \epsilon^\mathrm{new}(\omega) \equiv
    \epsilon^\mathrm{GGA}(\omega + E_g^\mathrm{new} - E_g^\mathrm{GGA}) \, .
\end{equation}
A total of 18 materials were excluded for unreliable band-gap calculations. Two were excluded because the implied shifts were negative, which would yield nonphysical results. As an additional precaution, we mask all sub-gap responses, setting
\begin{equation}
    \mathcal{W}(\omega < E_g, \textbf{q}) = 0\, .
\end{equation}
This step is essential, as sub-gap responses can arise in GGA computations due to numerical artifacts, and substantially affect the projected experimental sensitivity to light DM\@.
\end{enumerate}
In the following sections, we refer to this preprocessed dielectric tensor as $\tensor\epsilon_{\mathrm{data}}$.

\ssection{Isotropic case}

Neglecting anisotropy, our fitted dielectric function is determined by identifying the roots $\hat{\omega}_k$ of $\Re(\epsilon)$ at $q=0$ with component plasma frequencies, each of which is represented by a Lindhard dielectric function. The remaining parameters are then determined by direct numerical optimization in logarithmic coordinates. To implement this procedure effectively and efficiently, it is necessary to address the following complications:
\begin{enumerate}
    \item The widths and heights of plasmon peaks must be estimated a priori to seed the numerical optimization process.
    \item To account for noise in the data, a robust numerical definition of a root must be chosen to avoid misidentification of spurious features.
    \item Relatedly, when $\Re(\epsilon)$ is near zero, there may be several roots that correspond to the same physical feature due to numerical noise, and these must be grouped accordingly.
\end{enumerate}

\begin{figure*}\centering
    \includegraphics[width=\textwidth]{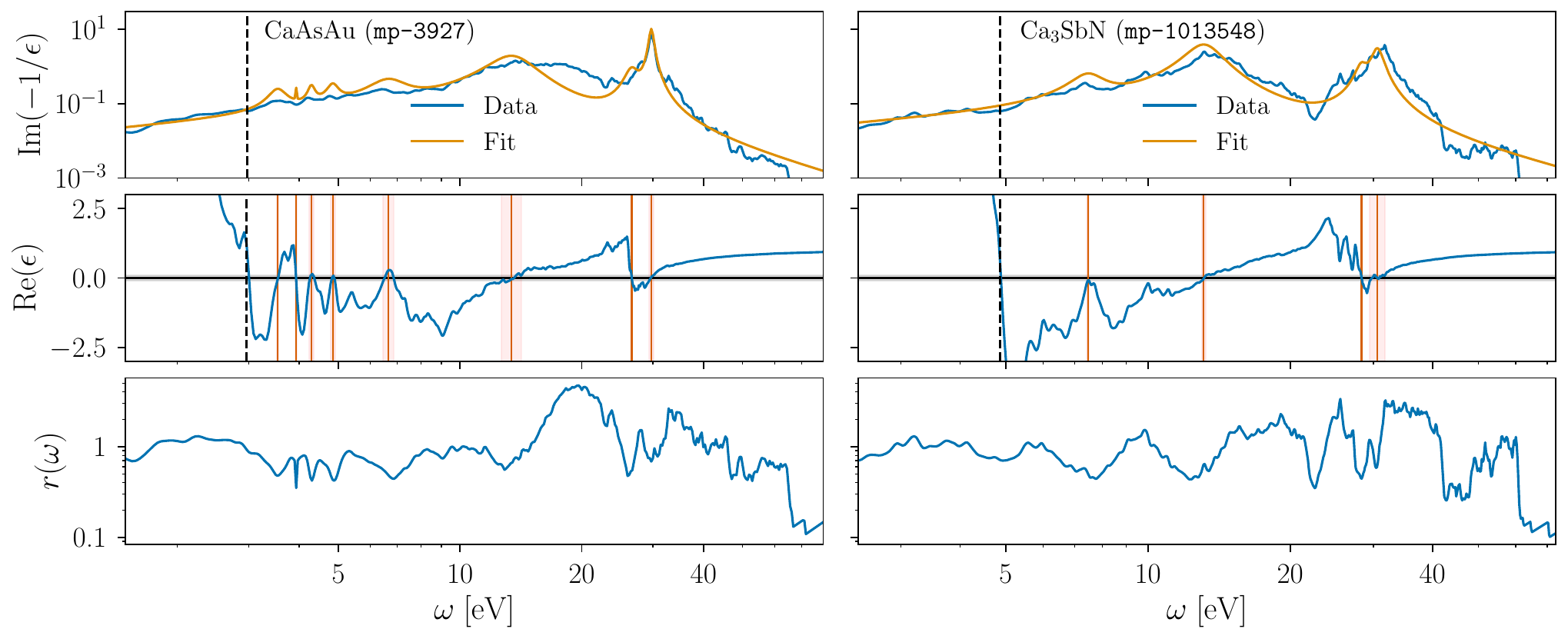}
    \caption{\textbf{Illustration of the fitting procedure for two example materials.} \textit{Top:} fitted loss functions $\mathcal W_{\fit}$ as defined in \cref{eq:combined-fit}.\ \textit{Middle:} identification of roots in the real part of the dielectric function. Roots after the first root are used as peak frequencies for fit components and are indicated with vertical lines. Red shaded regions indicate the zero-crossing regions used to estimate peak width for seeding numerical optimization.\ \textit{Bottom:} The ratio $r(\omega)$ between the loss function obtained directly from the data, $\mathcal W_{\mathrm{data}}$, and the loss function obtained from the fitted Lindhard functions, $\mathcal W_{\fit}$. We assume this ratio is independent of momentum.}
    \label{fig:sm-example-fits}
\end{figure*}

The estimation of peak heights is easy to address: they can be estimated by the value of the loss function in the data itself at each $\hat\omega_k$. The other issues require some additional structure. To avoid the misidentification of spurious roots from fluctuations, we define a threshold region $(-\tau, \tau)$ around $\Re(\epsilon) = 0$. A region $R$ in $\omega$ is determined to contain a root when the value of $\Re(\epsilon)$ crosses from $\left|\Re(\epsilon)\right| > \tau$ to $\left|\Re(\epsilon)\right| < \tau$ for at least two consecutive points, or when $\Re(\epsilon)$ crosses from one side of the threshold region to the other. To group nearby roots, we define a second threshold $\tau' = m\tau$ for a constant $m$. If $\left|\Re(\epsilon)\right|$ does not exceed $\tau'$ at any point between two consecutive root regions $R_1$ and $R_2$, then these are combined into a region $R = (\min R_1, \max R_2)$. We use $\tau = 0.1$ and $m = 3.5$, which eliminates most of the spurious roots associated with noise. Conveniently, partitioning $\omega$ into such regions also allows us to estimate the peak width as a constant multiple of the width of the region containing the corresponding root. We choose this constant so that the estimated peak width is twice the absolute value of the average derivative of $\Re(\epsilon)$ in the zero-crossing region. With these adjustments, our fitting procedure can be applied to every material in our dataset in just a few CPU-minutes. The identification of roots and their corresponding regions is illustrated for two example materials in \cref{fig:sm-example-fits}.

In order to assess the reliability of our results, we evaluate two quantitative benchmarks of the robustness of our fitting procedure. One benchmark is direct comparison between the data and our fits in the $q=0$ limit. Recall that we define $\mathcal{W}_{\fit}$ in a form that does not reduce to $\mathcal{W}_{\mathrm{data}}$ in this limit, and we absorb the residual in the factor $r(\omega)$. A key assumption is that $r(\omega)$ has at most mild $q$ dependence, and this is most easily justifiable if $r(\omega)$ is generally $\mathcal O(1)$ at small $q$. The value of $r(\omega) - 1$ quantifies the failure of our Lindhard fit to describe the response at small $q$, so if $r(\omega) - 1$ is large, there is no reason to expect that the $q$-dependence of the Lindhard function is a reliable proxy for the $q$-dependence of the dielectric function in question.

To quantify the goodness of fit, we use an analogue of the reduced $\chi^2$ statistic in logarithmic coordinates. We define
\begin{equation}
    \label{eq:sm-fit-metric}
    \left\langle\xi^2\right\rangle
        \left[\mathcal W_{\fit}, \mathcal W_{\mathrm{data}}\right]
    \equiv
    \frac{1}{N}\sum_{k=1}^{N}\,\biggl(
        \log_{10}\left[\left.\mathcal W_{\fit}(\omega_k)\right|_{q=0}\right]
        -
        \log_{10}\left[\mathcal W_{\mathrm{data}}(\omega_k)\right]
    \biggr)^2
    ,
\end{equation}
where $\mathcal W_{\mathrm{data}}$ denotes the loss function $\Im(-1/\epsilon)$ derived from data, $\mathcal W_{\fit}$ is the loss function as defined in \cref{eq:combined-fit}, and the $\omega_k$ are $N$ geometrically-spaced points above $\hat\omega_0$. Note that $\mathcal W_{\fit}$ is the loss function corresponding directly to the combination of Lindhard dielectric functions. Our actual computations are based on $\mathcal W_r$ from \cref{eq:fit-with-ratio}, which agrees exactly with $\mathcal W_{\mathrm{data}}$ in the limit $\bb q \to 0$. Thus, this comparison tests the goodness of fit of the combination of Lindhard functions itself, before the factor $r(\omega)$ is included.

The second benchmark is based on the two sum rules satisfied by the loss function:
\begin{equation}
    \label{eq:sm-sum-rules}
    \int_0^\infty\du\omega\, \omega \mathcal{W} = \tfrac12\pi\omega_\plas^2
    ,
    \qquad
    \lim_{q \,\rightarrow \,0}\int_0^\infty\du\omega\,
        \frac{\mathcal{W}}{\omega} = \tfrac12\pi
    .
\end{equation}
We define $\Lambda = \int_0^\infty\du\omega\, \omega \mathcal{W}$ and $\lambda = \lim_{q \,\rightarrow \,0}\int_0^\infty\du\omega\, \omega^{-1}\mathcal{W}$, and anticipate that $\Lambda/(\frac12\pi\omega_\plas^2)=\lambda/(\frac\pi2)=1$ for a perfectly-modeled material. In practice, $\Lambda$ is challenging to estimate: the corresponding sum rule originates from charge conservation, and the plasma frequency $\omega_\plas$ is used as a proxy for the density of charge carriers. This is true of the Lindhard model, for example, but since a typical material in our dataset is fit with several Lindhard components simultaneously, a unique $\omega_\plas$ cannot be readily extracted. Thus, we only evaluate $\Lambda$ for materials which have one plasmon, and are accordingly fitted with a single Lindhard function, a subset of 122 materials.

\begin{figure*}\centering
    \includegraphics[width=\textwidth]{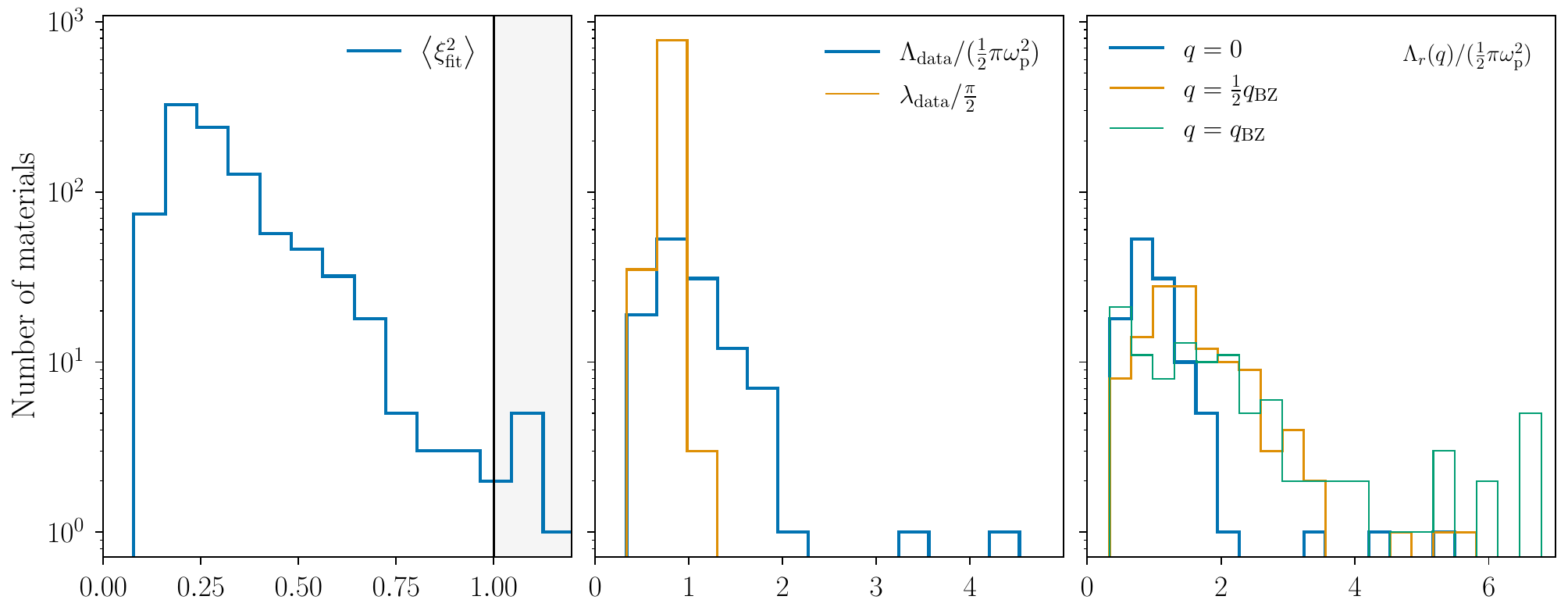}
    \caption{\textbf{Goodness-of-fit metrics for directionally-averaged fits.}
    \ \textit{Left:} logarithmic average distance $\langle\xi^2\rangle$ (\cref{eq:sm-fit-metric}) between fit and data. Materials lying in the shaded region are excluded from consideration.
    \ \textit{Center:} test of the sum rules in \cref{eq:sm-sum-rules} in the dataset. When the sum rules are satisfied, $\Lambda/(\frac12\pi\omega_\plas^2) = \lambda/\frac\pi2 = 1$.
    $\lambda_\mathrm{data}$ includes all materials used in the dataset while $\Lambda_\mathrm{data}$ includes 127 materials with a single plasmon peak. Here $\Lambda_{\mathrm{data}}$ is evaluated at $q=0$, corresponding to the available data.
    \ \textit{Right:} distribution of $\Lambda$ for the loss function $\mathcal W_r$ (\cref{eq:fit-with-ratio}) for several momenta for the same set of 127 materials, in units of the size of the Brillouin zone, $q_{\mathrm{BZ}}$. At $q=0$ (blue curve), $\mathcal W_r = \mathcal W_{\mathrm{data}}$ by construction, so $\Lambda_r = \Lambda_{\mathrm{data}}$.}
    \label{fig:sm-sum-rules}
\end{figure*}

\Cref{fig:sm-sum-rules} shows the combination of these benchmarks. The left panel shows the distribution of $\langle\xi^2\rangle$ between the data and $\mathcal W_{\fit}$. A value of $\langle\xi^2\rangle = 1$ would correspond to an average discrepancy of a factor of $10$, so the strong peak at $\langle\xi^2\rangle \sim 0.1$ indicates that our fits are generally quite accurate at the $\mathcal O(1)$ level before the ratio factor $r(\omega)$ is included. We conservatively exclude all materials with $\langle\xi^2\rangle > 1$, indicated by the shaded region. The middle panel shows the values of $\Lambda$ and $\lambda$ at $q=0$ in the data itself. The orange curve shows the distribution of $\lambda/(\frac12\pi)$, which is purely a consistency test of the data itself: to the extent that the dielectric responses in the \mpr{} dataset are physically accurate, they should satisfy $\lambda/(\frac12\pi) = 1$. The blue curve shows the distribution of $\Lambda/(\frac12\pi\omega_\plas^2)$, which is a measure of uncertainty in both the data and the fit, since both are used to determine $\omega_\plas$.

The right panel shows the values of $\Lambda$ corresponding to our actual loss functions, $\mathcal W_r$, for several values of $q$. We denote these values by $\Lambda_r$. At $q=0$, $\mathcal W_r = W_{\mathrm{data}}$ by construction, so the corresponding distribution (blue) exactly matches the distribution in the data (middle panel). At higher momenta, when $q$ becomes comparable to the size $q_{\mathrm{BZ}}$ of the Brillouin zone, the discrepancy between $\Lambda$ and $\frac12\pi\omega_\plas^2$ becomes noticeably larger. However, we stress that the vertical axis is logarithmic, and only a small handful of materials exhibit $\Lambda_r / (\frac12\pi\omega_\plas^2) > 3$.

\ssection{Anisotropic case}

\ssubsection{Computation of daily modulation}

Before assessing the daily modulation, we identify the anisotropic materials in the dataset, which are characterized by loss functions that vary between different axes. To quantify the degree of anisotropy, we use the metric $\langle\xi^2\rangle$ between loss functions in different axes, which we denote $\langle\xi_{\mathrm{ax},ij}^2\rangle$ for axes $i$ and $j$.

\begin{figure*}\centering
    \includegraphics[width=\textwidth]{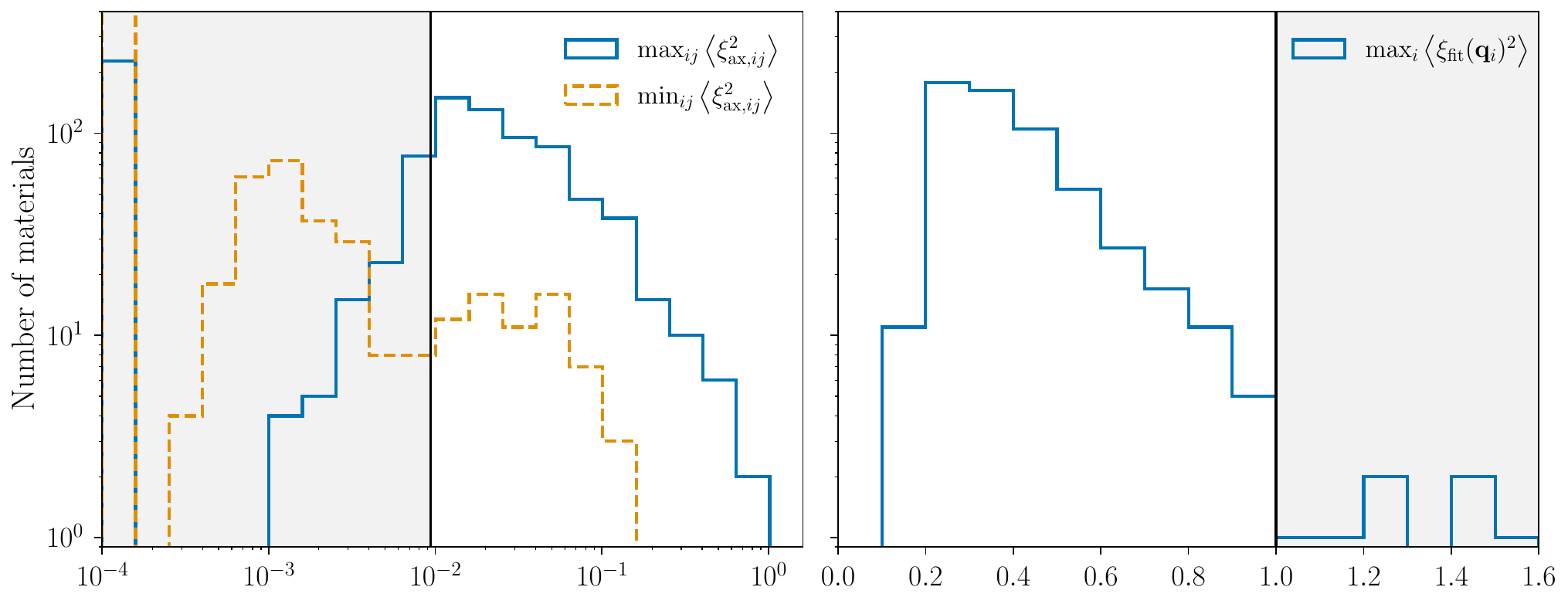}
    \caption{\textbf{Summary of anisotropic fits.}
    \ \textit{Left:} distribution of anisotropy in the data, as measured by $\langle\xi^2_{\mathrm{ax},ij}\rangle$. Materials with a large maximum anisotropy and a low minimum anisotropy are interpreted as having planar symmetry. 
    The gray shaded region indicates the threshold applied: materials in the blue histogram falling below this threshold are classified as isotropic, while those in the orange histogram are identified as possessing planar symmetry.
    \ \textit{Right:} maximum value of the goodness-of-fit statistic (\cref{eq:sm-fit-metric}) across all axes for each fit performed for an anisotropic material. Materials in the gray shaded region were excluded.}
    \label{fig:sm-anisotropic-fit}
\end{figure*}

Due to the sensitivity of the fitting procedure to noise in the data, we only consider a material to be anisotropic at all if the loss function differs by more than 25\% between at least one pair of axes. This corresponds to the condition $\max_{ij} \langle\xi_{\mathrm{ax},ij}^2\rangle \gtrsim 9.4 \times 10^{-3}$. The distribution of this quantity is shown by the blue curve in the left panel of \cref{fig:sm-anisotropic-fit}. In total, we find that \num{593} materials are anisotropic according to our selection criteria. Of these, \num{527} have a planar symmetry. 

We also assess the quality of our fitting procedure by computing the maximum value of $\langle\xi^2_{\fit}(\hat{\bb q}_i) \rangle$ across all \num{800} directions for which we fit the loss function. These maxima are shown in the right panel of \cref{fig:sm-anisotropic-fit}. We find that by this metric, the fits are on average within an order of magnitude of the data. However, this should be interpreted with caution, as the calculated sensitivity of some materials to DM depends on the quality of fitting to distinct features such as plasmons, which is not necessarily quantified by this single number. We find that \num{16} materials are incompatible with our fitting procedure, and \num{7} exhibit $\langle\xi^2_{\fit}(\hat{\bb q}_i)\rangle > 1$ as  can be seen in the right panel of \cref{fig:sm-anisotropic-fit}. These materials are excluded, leaving \num{570} materials to be considered in the anisotropic scattering analysis.

We compute the rate according to \cref{eq:rate} over a 24 hour interval, utilizing the \texttt{vegasflow} package~\cite{Carrazza:2020rdn, vegasflow_package} to carry out integration, using \num{e7} integration points and three iterations. To reduce computational expense, for materials with planar symmetry, we only compute the rate over a 12 hour interval, since the modulation over the other 12 hours can be obtained by a reflection symmetry.

To assess convergence, we compute the rate in the first hour twice, using the discrepancy between the two independently computed values as an estimate of the relative error in the computation. Due to the variety of scales incorporated in the integration, and sensitivity to the initial random seed, we find this yields a more reliable estimate than the default integration error estimate of the VEGAS algorithm. As a convergence criterion, we require the relative discrepancy between these two successive computations to be less than 10\% of the peak-to-peak modulation, and we find that 95\% of cases exhibit discrepancies below 0.1\%. Since our statistical test averages over multiple hours, the effect of the residual errors is suppressed.

To benchmark our anisotropic fitting procedure, we evaluate our pipeline using the anisotropic electron gas (AEG) model of \refcite{Boyd:2022tcn}. The AEG model incorporates anisotropy via an anisotropic dispersion relation of the form
\begin{equation}
    E_{\bb q} = \frac{q_x^2 + q_y^2}{2 m_{xy}^2} + \frac{q_z^2}{2 m_z^2}
    ,
\end{equation}
where $\bb q$ denotes the momentum of an electron quasiparticle, and   $m_{xy}$ and $m_z$ represent effective masses delineating different responses in the $xy$ plane and the $z$ direction. The momentum-dependent dielectric tensor $\tensor\epsilon(\bb q, \omega)$ can be computed explicitly in the random phase approximation, giving an anisotropic analogue of the Lindhard dielectric function. Our fit is based only on the zero-momentum dielectric tensor, $\left.\tensor\epsilon(\omega)\right|_{\bb q=0}$. We can thus test our pipeline as follows: we choose a set of AEG model parameters; sample $\left.\tensor\epsilon(\omega)\right|_{\bb q=0}$; fit to these samples; and then compare the rate modulation implied by our fit to that computed directly from the AEG\@.

\begin{figure*}\centering
    \includegraphics[width=\textwidth]{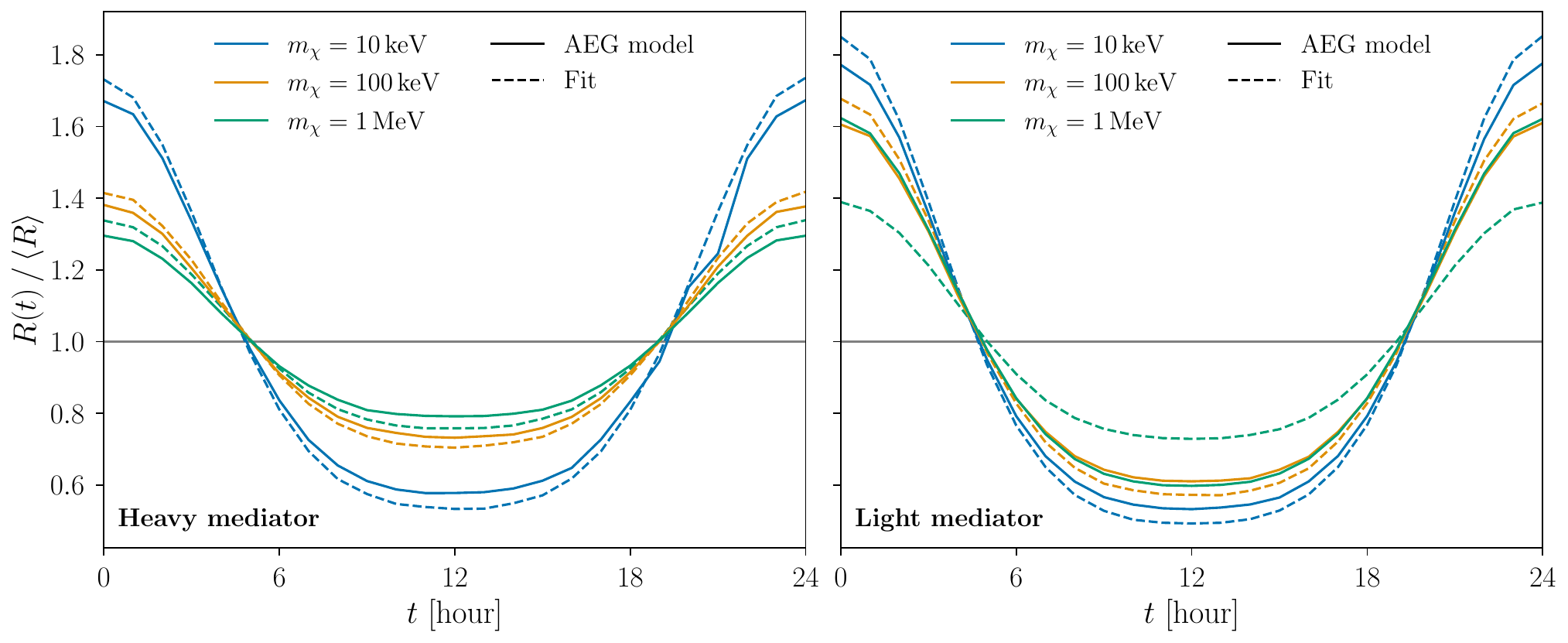}
    \caption{\textbf{Daily modulation.} Daily modulation of the DM scattering rate for an anisotropic electron gas model with $m_{xy} = \qty{370}{\kilo\electronvolt}$, $m_z = \qty{7.37}{\mega\electronvolt}$, $\omega_\plas = \qty{8.76}{\electronvolt}$, and $\Gamma_\plas = \qty{876}{\milli\electronvolt}$. The solid curves delineate the ground truth whereas the dashed curves delineate the results of our fitting procedure.}
    \label{fig:sm-aeg-benchmark}
\end{figure*}

The results of this test are summarized in \cref{fig:sm-aeg-benchmark}. The amplitude of the daily modulation for the fit agrees with the original model to within an $\mathcal O(1)$ factor. The differences are attributable to the different momentum dependence (\textit{i.e.} dispersion relation) between the AEG model and our fit. In general, the anisotropic momentum dependence is expected to be material-dependent: materials where anisotropy is mainly driven by the dispersion relation for electronic excitations would be better modeled by the AEG, and otherwise might be better modeled by our fitting function, \textit{i.e.}, \cref{eq:combined-fit} of the main text.

\ssubsection{Statistical detection of directionality}
In this work, we use a simple criterion for sensitivity to a directional DM signal. We consider the directionality to be detectable at a certain cross section if, given the implied number of events, there is a 95\% chance that the hypothesis of an isotropic signal can be rejected at the 95\% confidence level (CL).

To compute this number, we assume that the number $N$ of scattering events in a fixed exposure $T$ is Poisson distributed, \textit{i.e.}, $N \sim \operatorname{Poisson}[RT]$, where $R$ is the average scattering rate (\cref{eq:rate}). We then consider the daily modulation due to anisotropic sensitivity with the same parameters as those chosen in \refcite{Boyd:2022tcn}, and we divide each day into two half-days with scattering rates $R_{\mathrm{AM}}$ and $R_{\mathrm{PM}}$.

Now we wish to reject the hypothesis that $R_{\mathrm{AM}} = R_{\mathrm{PM}}$. The difference between the event counts in each of the half-day bins is Skellam-distributed, \textit{i.e.}, $\Delta N \equiv N_{\mathrm{AM}} - N_{\mathrm{PM}} \sim \operatorname{Skellam}[R_{\mathrm{AM}}, R_{\mathrm{PM}}]$. The pdf of this distribution is given by
\begin{equation}
    f_{\mathrm{S}}[R_{\mathrm{AM}}T, R_{\mathrm{PM}}T](\Delta N) =
    e^{-(R_{\mathrm{AM}}T + R_{\mathrm{PM}}T)}
    \left(\frac{R_{\mathrm{AM}}}{R_{\mathrm{PM}}}\right)^{\Delta N/2}
    I_{\Delta N}\left(2 \sqrt{(R_{\mathrm{AM}}T) (R_{\mathrm{PM}}T)}\right)
    ,
\end{equation}
where $I$ denotes the modified Bessel function of the first kind. Given this pdf, and fixing the exposure $T$, we determine the difference $\Delta N$ that corresponds to rejection of the hypothesis that $R_{\mathrm{AM}} = R_{\mathrm{PM}}$ at 95\% CL\@. We then use Monte Carlo sampling to identify the total number of events, $N_{\mathrm{AM}} + N_{\mathrm{PM}}$, such that the hypothesis is rejected in 95\% of cases. This total number of events in turn corresponds to a minimum exposure required to detect the anisotropy of the DM signal.

Note that this method gives conservative estimates for the directional detection reach. Given a DM halo model, one can use the entire shape of the rate modulation curve to test for directionality, using \textit{e.g.} optimal filtering. Thus, we expect that a directional DM signal can be identified at smaller cross sections than indicated in our results.

\ssection{Summary of additional results}

In the main text, we show reach curves for absorption and for scattering via a light mediator, each for a small subset of materials that are optimal in the \mpr{} dataset according to our criteria. Here, we supply results for the two remaining cases: scattering via a heavy mediator and scattering or absorption with all of the other materials in our dataset.

\begin{figure*}\centering
    \includegraphics[width=\textwidth]{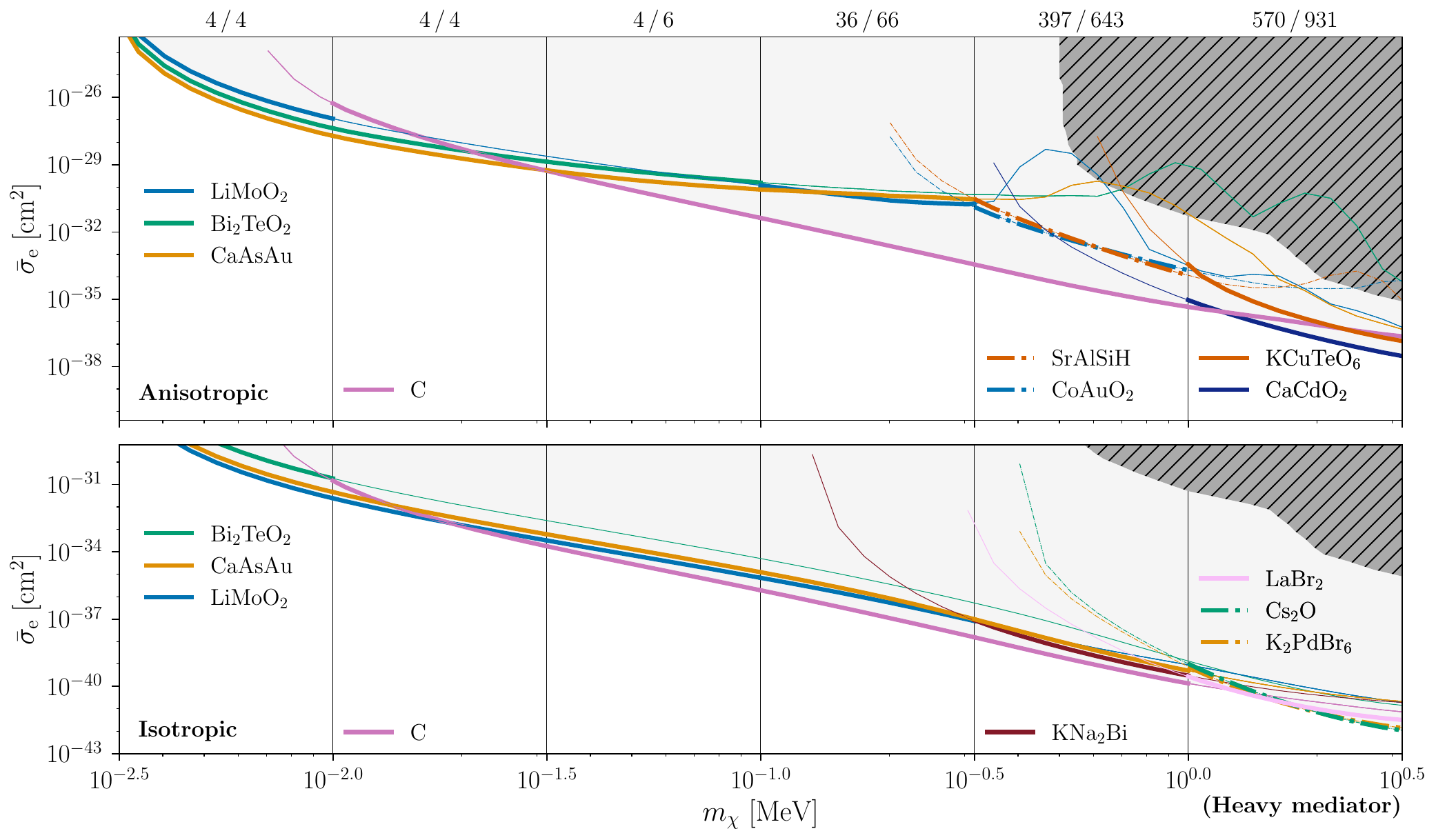}
    \caption{\textbf{Scattering via heavy mediator.} Projected reach of optimal materials for DM scattering via a heavy mediator. Existing constraints from \refscite{Barak:2020fql,Amaral:2020ryn,Aguilar-Arevalo:2019wdi,Essig:2017kqs,Agnes:2018oej,Aprile:2019xxb} are indicated by the hatched dark gray region. All other features are identical to \cref{fig:scattering-light}.}
    \label{fig:scattering-heavy}
\end{figure*}

First, we present our results for DM scattering via a heavy mediator in \cref{fig:scattering-heavy}. The hatched region in \cref{fig:scattering-heavy} indicates existing constraints on the relevant parameter space. As mentioned in the main text, the qualitative outcome of this analysis is very similar to the case of the light mediator that we show in \cref{fig:scattering-light}. The full set of optimal materials is very similar between the two cases.

Next, for all the cases for which projections appear in this work, and for all other materials in the \materialsproject{} dataset, we have made our data products publicly available~\cite{full_repository} to facilitate future studies. For each material in our dataset, the repository includes the following properties:
\begin{itemize}
    \item The full dielectric tensor data in the limit $q\to 0$;
    \item Other material data such as band gaps and densities;
    \item Projected reach curves for absorption, isotropic scattering, and anisotropic scattering;
    \item Modulation curves for anisotropic materials, error estimates, and planar symmetries.
\end{itemize}
Dielectric response data in the repository is shifted per the procedure described in \cref{eq:gap-shift}. Further documentation is available in the repository itself, and usage examples are included in the form of a Jupyter notebook included within the repository.

\bibliography{references}

%apsrev4-2.bst 2019-01-14 (MD) hand-edited version of apsrev4-1.bst
%Control: key (0)
%Control: author (8) initials jnrlst
%Control: editor formatted (1) identically to author
%Control: production of article title (0) allowed
%Control: page (0) single
%Control: year (1) truncated
%Control: production of eprint (0) enabled
\begin{thebibliography}{88}%
\makeatletter
\providecommand \@ifxundefined [1]{%
 \@ifx{#1\undefined}
}%
\providecommand \@ifnum [1]{%
 \ifnum #1\expandafter \@firstoftwo
 \else \expandafter \@secondoftwo
 \fi
}%
\providecommand \@ifx [1]{%
 \ifx #1\expandafter \@firstoftwo
 \else \expandafter \@secondoftwo
 \fi
}%
\providecommand \natexlab [1]{#1}%
\providecommand \enquote  [1]{``#1''}%
\providecommand \bibnamefont  [1]{#1}%
\providecommand \bibfnamefont [1]{#1}%
\providecommand \citenamefont [1]{#1}%
\providecommand \href@noop [0]{\@secondoftwo}%
\providecommand \href [0]{\begingroup \@sanitize@url \@href}%
\providecommand \@href[1]{\@@startlink{#1}\@@href}%
\providecommand \@@href[1]{\endgroup#1\@@endlink}%
\providecommand \@sanitize@url [0]{\catcode `\\12\catcode `\$12\catcode `\&12\catcode `\#12\catcode `\^12\catcode `\_12\catcode `\%12\relax}%
\providecommand \@@startlink[1]{}%
\providecommand \@@endlink[0]{}%
\providecommand \url  [0]{\begingroup\@sanitize@url \@url }%
\providecommand \@url [1]{\endgroup\@href {#1}{\urlprefix }}%
\providecommand \urlprefix  [0]{URL }%
\providecommand \Eprint [0]{\href }%
\providecommand \doibase [0]{https://doi.org/}%
\providecommand \selectlanguage [0]{\@gobble}%
\providecommand \bibinfo  [0]{\@secondoftwo}%
\providecommand \bibfield  [0]{\@secondoftwo}%
\providecommand \translation [1]{[#1]}%
\providecommand \BibitemOpen [0]{}%
\providecommand \bibitemStop [0]{}%
\providecommand \bibitemNoStop [0]{.\EOS\space}%
\providecommand \EOS [0]{\spacefactor3000\relax}%
\providecommand \BibitemShut  [1]{\csname bibitem#1\endcsname}%
\let\auto@bib@innerbib\@empty
%</preamble>
\bibitem [{\citenamefont {Essig}\ \emph {et~al.}(2012)\citenamefont {Essig}, \citenamefont {Mardon},\ and\ \citenamefont {Volansky}}]{Essig:2011nj}%
  \BibitemOpen
  \bibfield  {author} {\bibinfo {author} {\bibfnamefont {R.}~\bibnamefont {Essig}}, \bibinfo {author} {\bibfnamefont {J.}~\bibnamefont {Mardon}},\ and\ \bibinfo {author} {\bibfnamefont {T.}~\bibnamefont {Volansky}},\ }\bibfield  {title} {\bibinfo {title} {{Direct Detection of Sub-GeV Dark Matter}},\ }\href {https://doi.org/10.1103/PhysRevD.85.076007} {\bibfield  {journal} {\bibinfo  {journal} {Phys. Rev. D}\ }\textbf {\bibinfo {volume} {85}},\ \bibinfo {pages} {076007} (\bibinfo {year} {2012})},\ \Eprint {https://arxiv.org/abs/1108.5383} {arXiv:1108.5383 [hep-ph]} \BibitemShut {NoStop}%
\bibitem [{\citenamefont {Derenzo}\ \emph {et~al.}(2017)\citenamefont {Derenzo}, \citenamefont {Essig}, \citenamefont {Massari}, \citenamefont {Soto},\ and\ \citenamefont {Yu}}]{Derenzo:2016fse}%
  \BibitemOpen
  \bibfield  {author} {\bibinfo {author} {\bibfnamefont {S.}~\bibnamefont {Derenzo}}, \bibinfo {author} {\bibfnamefont {R.}~\bibnamefont {Essig}}, \bibinfo {author} {\bibfnamefont {A.}~\bibnamefont {Massari}}, \bibinfo {author} {\bibfnamefont {A.}~\bibnamefont {Soto}},\ and\ \bibinfo {author} {\bibfnamefont {T.-T.}\ \bibnamefont {Yu}},\ }\bibfield  {title} {\bibinfo {title} {{Direct Detection of sub-GeV Dark Matter with Scintillating Targets}},\ }\href {https://doi.org/10.1103/PhysRevD.96.016026} {\bibfield  {journal} {\bibinfo  {journal} {Phys. Rev. D}\ }\textbf {\bibinfo {volume} {96}},\ \bibinfo {pages} {016026} (\bibinfo {year} {2017})},\ \Eprint {https://arxiv.org/abs/1607.01009} {arXiv:1607.01009 [hep-ph]} \BibitemShut {NoStop}%
\bibitem [{\citenamefont {Essig}\ \emph {et~al.}(2016)\citenamefont {Essig}, \citenamefont {Fernandez-Serra}, \citenamefont {Mardon}, \citenamefont {Soto}, \citenamefont {Volansky},\ and\ \citenamefont {Yu}}]{Essig:2015cda}%
  \BibitemOpen
  \bibfield  {author} {\bibinfo {author} {\bibfnamefont {R.}~\bibnamefont {Essig}}, \bibinfo {author} {\bibfnamefont {M.}~\bibnamefont {Fernandez-Serra}}, \bibinfo {author} {\bibfnamefont {J.}~\bibnamefont {Mardon}}, \bibinfo {author} {\bibfnamefont {A.}~\bibnamefont {Soto}}, \bibinfo {author} {\bibfnamefont {T.}~\bibnamefont {Volansky}},\ and\ \bibinfo {author} {\bibfnamefont {T.-T.}\ \bibnamefont {Yu}},\ }\bibfield  {title} {\bibinfo {title} {{Direct Detection of sub-GeV Dark Matter with Semiconductor Targets}},\ }\href {https://doi.org/10.1007/JHEP05(2016)046} {\bibfield  {journal} {\bibinfo  {journal} {JHEP}\ }\textbf {\bibinfo {volume} {05}},\ \bibinfo {pages} {046}},\ \Eprint {https://arxiv.org/abs/1509.01598} {arXiv:1509.01598 [hep-ph]} \BibitemShut {NoStop}%
\bibitem [{\citenamefont {Graham}\ \emph {et~al.}(2012)\citenamefont {Graham}, \citenamefont {Kaplan}, \citenamefont {Rajendran},\ and\ \citenamefont {Walters}}]{Graham:2012su}%
  \BibitemOpen
  \bibfield  {author} {\bibinfo {author} {\bibfnamefont {P.~W.}\ \bibnamefont {Graham}}, \bibinfo {author} {\bibfnamefont {D.~E.}\ \bibnamefont {Kaplan}}, \bibinfo {author} {\bibfnamefont {S.}~\bibnamefont {Rajendran}},\ and\ \bibinfo {author} {\bibfnamefont {M.~T.}\ \bibnamefont {Walters}},\ }\bibfield  {title} {\bibinfo {title} {{Semiconductor Probes of Light Dark Matter}},\ }\href {https://doi.org/10.1016/j.dark.2012.09.001} {\bibfield  {journal} {\bibinfo  {journal} {Phys. Dark Univ.}\ }\textbf {\bibinfo {volume} {1}},\ \bibinfo {pages} {32} (\bibinfo {year} {2012})},\ \Eprint {https://arxiv.org/abs/1203.2531} {arXiv:1203.2531 [hep-ph]} \BibitemShut {NoStop}%
\bibitem [{\citenamefont {Hochberg}\ \emph {et~al.}(2017{\natexlab{a}})\citenamefont {Hochberg}, \citenamefont {Lin},\ and\ \citenamefont {Zurek}}]{Hochberg:2016sqx}%
  \BibitemOpen
  \bibfield  {author} {\bibinfo {author} {\bibfnamefont {Y.}~\bibnamefont {Hochberg}}, \bibinfo {author} {\bibfnamefont {T.}~\bibnamefont {Lin}},\ and\ \bibinfo {author} {\bibfnamefont {K.~M.}\ \bibnamefont {Zurek}},\ }\bibfield  {title} {\bibinfo {title} {{Absorption of light dark matter in semiconductors}},\ }\href {https://doi.org/10.1103/PhysRevD.95.023013} {\bibfield  {journal} {\bibinfo  {journal} {Phys. Rev.}\ }\textbf {\bibinfo {volume} {D95}},\ \bibinfo {pages} {023013} (\bibinfo {year} {2017}{\natexlab{a}})},\ \Eprint {https://arxiv.org/abs/1608.01994} {arXiv:1608.01994 [hep-ph]} \BibitemShut {NoStop}%
%%CITATION = ARXIV:1608.01994;%%
\bibitem [{\citenamefont {Griffin}\ \emph {et~al.}(2021)\citenamefont {Griffin}, \citenamefont {Hochberg}, \citenamefont {Inzani}, \citenamefont {Kurinsky}, \citenamefont {Lin},\ and\ \citenamefont {Yu}}]{Griffin:2020lgd}%
  \BibitemOpen
  \bibfield  {author} {\bibinfo {author} {\bibfnamefont {S.~M.}\ \bibnamefont {Griffin}}, \bibinfo {author} {\bibfnamefont {Y.}~\bibnamefont {Hochberg}}, \bibinfo {author} {\bibfnamefont {K.}~\bibnamefont {Inzani}}, \bibinfo {author} {\bibfnamefont {N.}~\bibnamefont {Kurinsky}}, \bibinfo {author} {\bibfnamefont {T.}~\bibnamefont {Lin}},\ and\ \bibinfo {author} {\bibfnamefont {T.~C.}\ \bibnamefont {Yu}},\ }\bibfield  {title} {\bibinfo {title} {{Silicon carbide detectors for sub-GeV dark matter}},\ }\href {https://doi.org/10.1103/PhysRevD.103.075002} {\bibfield  {journal} {\bibinfo  {journal} {Phys. Rev. D}\ }\textbf {\bibinfo {volume} {103}},\ \bibinfo {pages} {075002} (\bibinfo {year} {2021})},\ \Eprint {https://arxiv.org/abs/2008.08560} {arXiv:2008.08560 [hep-ph]} \BibitemShut {NoStop}%
\bibitem [{\citenamefont {Hochberg}\ \emph {et~al.}(2016)\citenamefont {Hochberg}, \citenamefont {Zhao},\ and\ \citenamefont {Zurek}}]{Hochberg:2015pha}%
  \BibitemOpen
  \bibfield  {author} {\bibinfo {author} {\bibfnamefont {Y.}~\bibnamefont {Hochberg}}, \bibinfo {author} {\bibfnamefont {Y.}~\bibnamefont {Zhao}},\ and\ \bibinfo {author} {\bibfnamefont {K.~M.}\ \bibnamefont {Zurek}},\ }\bibfield  {title} {\bibinfo {title} {{Superconducting Detectors for Superlight Dark Matter}},\ }\href {https://doi.org/10.1103/PhysRevLett.116.011301} {\bibfield  {journal} {\bibinfo  {journal} {Phys. Rev. Lett.}\ }\textbf {\bibinfo {volume} {116}},\ \bibinfo {pages} {011301} (\bibinfo {year} {2016})},\ \Eprint {https://arxiv.org/abs/1504.07237} {arXiv:1504.07237 [hep-ph]} \BibitemShut {NoStop}%
\bibitem [{\citenamefont {Hochberg}\ \emph {et~al.}(2019)\citenamefont {Hochberg}, \citenamefont {Charaev}, \citenamefont {Nam}, \citenamefont {Verma}, \citenamefont {Colangelo},\ and\ \citenamefont {Berggren}}]{Hochberg:2019cyy}%
  \BibitemOpen
  \bibfield  {author} {\bibinfo {author} {\bibfnamefont {Y.}~\bibnamefont {Hochberg}}, \bibinfo {author} {\bibfnamefont {I.}~\bibnamefont {Charaev}}, \bibinfo {author} {\bibfnamefont {S.-W.}\ \bibnamefont {Nam}}, \bibinfo {author} {\bibfnamefont {V.}~\bibnamefont {Verma}}, \bibinfo {author} {\bibfnamefont {M.}~\bibnamefont {Colangelo}},\ and\ \bibinfo {author} {\bibfnamefont {K.~K.}\ \bibnamefont {Berggren}},\ }\bibfield  {title} {\bibinfo {title} {{Detecting Sub-GeV Dark Matter with Superconducting Nanowires}},\ }\href {https://doi.org/10.1103/PhysRevLett.123.151802} {\bibfield  {journal} {\bibinfo  {journal} {Phys. Rev. Lett.}\ }\textbf {\bibinfo {volume} {123}},\ \bibinfo {pages} {151802} (\bibinfo {year} {2019})},\ \Eprint {https://arxiv.org/abs/1903.05101} {arXiv:1903.05101 [hep-ph]} \BibitemShut {NoStop}%
\bibitem [{\citenamefont {Hochberg}\ \emph {et~al.}(2022)\citenamefont {Hochberg}, \citenamefont {Lehmann}, \citenamefont {Charaev}, \citenamefont {Chiles}, \citenamefont {Colangelo}, \citenamefont {Nam},\ and\ \citenamefont {Berggren}}]{Hochberg:2021yud}%
  \BibitemOpen
  \bibfield  {author} {\bibinfo {author} {\bibfnamefont {Y.}~\bibnamefont {Hochberg}}, \bibinfo {author} {\bibfnamefont {B.~V.}\ \bibnamefont {Lehmann}}, \bibinfo {author} {\bibfnamefont {I.}~\bibnamefont {Charaev}}, \bibinfo {author} {\bibfnamefont {J.}~\bibnamefont {Chiles}}, \bibinfo {author} {\bibfnamefont {M.}~\bibnamefont {Colangelo}}, \bibinfo {author} {\bibfnamefont {S.~W.}\ \bibnamefont {Nam}},\ and\ \bibinfo {author} {\bibfnamefont {K.~K.}\ \bibnamefont {Berggren}},\ }\bibfield  {title} {\bibinfo {title} {{New constraints on dark matter from superconducting nanowires}},\ }\href {https://doi.org/10.1103/PhysRevD.106.112005} {\bibfield  {journal} {\bibinfo  {journal} {Phys. Rev. D}\ }\textbf {\bibinfo {volume} {106}},\ \bibinfo {pages} {112005} (\bibinfo {year} {2022})},\ \Eprint {https://arxiv.org/abs/2110.01586} {arXiv:2110.01586 [hep-ph]} \BibitemShut {NoStop}%
\bibitem [{\citenamefont {Gao}\ \emph {et~al.}(2024)\citenamefont {Gao}, \citenamefont {Hochberg}, \citenamefont {Lehmann}, \citenamefont {Nam}, \citenamefont {Szypryt}, \citenamefont {Vissers},\ and\ \citenamefont {Xu}}]{Gao:2024irf}%
  \BibitemOpen
  \bibfield  {author} {\bibinfo {author} {\bibfnamefont {J.}~\bibnamefont {Gao}}, \bibinfo {author} {\bibfnamefont {Y.}~\bibnamefont {Hochberg}}, \bibinfo {author} {\bibfnamefont {B.~V.}\ \bibnamefont {Lehmann}}, \bibinfo {author} {\bibfnamefont {S.~W.}\ \bibnamefont {Nam}}, \bibinfo {author} {\bibfnamefont {P.}~\bibnamefont {Szypryt}}, \bibinfo {author} {\bibfnamefont {M.~R.}\ \bibnamefont {Vissers}},\ and\ \bibinfo {author} {\bibfnamefont {T.}~\bibnamefont {Xu}},\ }\href@noop {} {\bibinfo {title} {{Detecting Light Dark Matter with Kinetic Inductance Detectors}}} (\bibinfo {year} {2024}),\ \Eprint {https://arxiv.org/abs/2403.19739} {arXiv:2403.19739 [hep-ph]} \BibitemShut {NoStop}%
\bibitem [{\citenamefont {Schutz}\ and\ \citenamefont {Zurek}(2016)}]{Schutz:2016tid}%
  \BibitemOpen
  \bibfield  {author} {\bibinfo {author} {\bibfnamefont {K.}~\bibnamefont {Schutz}}\ and\ \bibinfo {author} {\bibfnamefont {K.~M.}\ \bibnamefont {Zurek}},\ }\bibfield  {title} {\bibinfo {title} {{Detectability of Light Dark Matter with Superfluid Helium}},\ }\href {https://doi.org/10.1103/PhysRevLett.117.121302} {\bibfield  {journal} {\bibinfo  {journal} {Phys. Rev. Lett.}\ }\textbf {\bibinfo {volume} {117}},\ \bibinfo {pages} {121302} (\bibinfo {year} {2016})},\ \Eprint {https://arxiv.org/abs/1604.08206} {arXiv:1604.08206 [hep-ph]} \BibitemShut {NoStop}%
%%CITATION = ARXIV:1604.08206;%%
\bibitem [{\citenamefont {Knapen}\ \emph {et~al.}(2017)\citenamefont {Knapen}, \citenamefont {Lin},\ and\ \citenamefont {Zurek}}]{Knapen:2016cue}%
  \BibitemOpen
  \bibfield  {author} {\bibinfo {author} {\bibfnamefont {S.}~\bibnamefont {Knapen}}, \bibinfo {author} {\bibfnamefont {T.}~\bibnamefont {Lin}},\ and\ \bibinfo {author} {\bibfnamefont {K.~M.}\ \bibnamefont {Zurek}},\ }\bibfield  {title} {\bibinfo {title} {{Light Dark Matter in Superfluid Helium: Detection with Multi-excitation Production}},\ }\href {https://doi.org/10.1103/PhysRevD.95.056019} {\bibfield  {journal} {\bibinfo  {journal} {Phys. Rev.}\ }\textbf {\bibinfo {volume} {D95}},\ \bibinfo {pages} {056019} (\bibinfo {year} {2017})},\ \Eprint {https://arxiv.org/abs/1611.06228} {arXiv:1611.06228 [hep-ph]} \BibitemShut {NoStop}%
%%CITATION = ARXIV:1611.06228;%%
\bibitem [{\citenamefont {Ashour}\ and\ \citenamefont {Griffin}(2024)}]{Ashour:2024xfp}%
  \BibitemOpen
  \bibfield  {author} {\bibinfo {author} {\bibfnamefont {O.~A.}\ \bibnamefont {Ashour}}\ and\ \bibinfo {author} {\bibfnamefont {S.~M.}\ \bibnamefont {Griffin}},\ }\bibfield  {title} {\bibinfo {title} {{Pressure-Tunable Targets for Light Dark Matter Direct Detection: The Case of Solid Helium}},\ }\href@noop {} {\  (\bibinfo {year} {2024})},\ \Eprint {https://arxiv.org/abs/2409.02439} {arXiv:2409.02439 [hep-ph]} \BibitemShut {NoStop}%
\bibitem [{\citenamefont {Hochberg}\ \emph {et~al.}(2017{\natexlab{b}})\citenamefont {Hochberg}, \citenamefont {Kahn}, \citenamefont {Lisanti}, \citenamefont {Tully},\ and\ \citenamefont {Zurek}}]{Hochberg:2016ntt}%
  \BibitemOpen
  \bibfield  {author} {\bibinfo {author} {\bibfnamefont {Y.}~\bibnamefont {Hochberg}}, \bibinfo {author} {\bibfnamefont {Y.}~\bibnamefont {Kahn}}, \bibinfo {author} {\bibfnamefont {M.}~\bibnamefont {Lisanti}}, \bibinfo {author} {\bibfnamefont {C.~G.}\ \bibnamefont {Tully}},\ and\ \bibinfo {author} {\bibfnamefont {K.~M.}\ \bibnamefont {Zurek}},\ }\bibfield  {title} {\bibinfo {title} {{Directional detection of dark matter with two-dimensional targets}},\ }\href {https://doi.org/10.1016/j.physletb.2017.06.051} {\bibfield  {journal} {\bibinfo  {journal} {Phys. Lett. B}\ }\textbf {\bibinfo {volume} {772}},\ \bibinfo {pages} {239} (\bibinfo {year} {2017}{\natexlab{b}})},\ \Eprint {https://arxiv.org/abs/1606.08849} {arXiv:1606.08849 [hep-ph]} \BibitemShut {NoStop}%
\bibitem [{\citenamefont {Cavoto}\ \emph {et~al.}(2018)\citenamefont {Cavoto}, \citenamefont {Luchetta},\ and\ \citenamefont {Polosa}}]{Cavoto:2017otc}%
  \BibitemOpen
  \bibfield  {author} {\bibinfo {author} {\bibfnamefont {G.}~\bibnamefont {Cavoto}}, \bibinfo {author} {\bibfnamefont {F.}~\bibnamefont {Luchetta}},\ and\ \bibinfo {author} {\bibfnamefont {A.~D.}\ \bibnamefont {Polosa}},\ }\bibfield  {title} {\bibinfo {title} {{Sub-GeV Dark Matter Detection with Electron Recoils in Carbon Nanotubes}},\ }\href {https://doi.org/10.1016/j.physletb.2017.11.064} {\bibfield  {journal} {\bibinfo  {journal} {Phys. Lett. B}\ }\textbf {\bibinfo {volume} {776}},\ \bibinfo {pages} {338} (\bibinfo {year} {2018})},\ \Eprint {https://arxiv.org/abs/1706.02487} {arXiv:1706.02487 [hep-ph]} \BibitemShut {NoStop}%
\bibitem [{\citenamefont {Blanco}\ \emph {et~al.}(2021)\citenamefont {Blanco}, \citenamefont {Kahn}, \citenamefont {Lillard},\ and\ \citenamefont {McDermott}}]{Blanco:2021hlm}%
  \BibitemOpen
  \bibfield  {author} {\bibinfo {author} {\bibfnamefont {C.}~\bibnamefont {Blanco}}, \bibinfo {author} {\bibfnamefont {Y.}~\bibnamefont {Kahn}}, \bibinfo {author} {\bibfnamefont {B.}~\bibnamefont {Lillard}},\ and\ \bibinfo {author} {\bibfnamefont {S.~D.}\ \bibnamefont {McDermott}},\ }\bibfield  {title} {\bibinfo {title} {{Dark Matter Daily Modulation With Anisotropic Organic Crystals}},\ }\href {https://doi.org/10.1103/PhysRevD.104.036011} {\bibfield  {journal} {\bibinfo  {journal} {Phys. Rev. D}\ }\textbf {\bibinfo {volume} {104}},\ \bibinfo {pages} {036011} (\bibinfo {year} {2021})},\ \Eprint {https://arxiv.org/abs/2103.08601} {arXiv:2103.08601 [hep-ph]} \BibitemShut {NoStop}%
\bibitem [{\citenamefont {Taufertsh\"ofer}\ \emph {et~al.}(2024)\citenamefont {Taufertsh\"ofer}, \citenamefont {Garcia-Sciveres},\ and\ \citenamefont {Griffin}}]{Taufertshofer:2023rgq}%
  \BibitemOpen
  \bibfield  {author} {\bibinfo {author} {\bibfnamefont {N.}~\bibnamefont {Taufertsh\"ofer}}, \bibinfo {author} {\bibfnamefont {M.}~\bibnamefont {Garcia-Sciveres}},\ and\ \bibinfo {author} {\bibfnamefont {S.~M.}\ \bibnamefont {Griffin}},\ }\bibfield  {title} {\bibinfo {title} {{Proposal for broad-range directional detection of light dark matter in cryogenic ice}},\ }\href {https://doi.org/10.1103/PhysRevD.110.103552} {\bibfield  {journal} {\bibinfo  {journal} {Phys. Rev. D}\ }\textbf {\bibinfo {volume} {110}},\ \bibinfo {pages} {103552} (\bibinfo {year} {2024})},\ \Eprint {https://arxiv.org/abs/2301.04778} {arXiv:2301.04778 [hep-ph]} \BibitemShut {NoStop}%
\bibitem [{\citenamefont {Adari}\ \emph {et~al.}(2023)\citenamefont {Adari} \emph {et~al.}}]{SENSEI:2023zdf}%
  \BibitemOpen
  \bibfield  {author} {\bibinfo {author} {\bibfnamefont {P.}~\bibnamefont {Adari}} \emph {et~al.} (\bibinfo {collaboration} {SENSEI}),\ }\href@noop {} {\bibinfo {title} {{SENSEI: First Direct-Detection Results on sub-GeV Dark Matter from SENSEI at SNOLAB}}} (\bibinfo {year} {2023}),\ \Eprint {https://arxiv.org/abs/2312.13342} {arXiv:2312.13342 [astro-ph.CO]} \BibitemShut {NoStop}%
\bibitem [{\citenamefont {Albakry}\ \emph {et~al.}(2024)\citenamefont {Albakry} \emph {et~al.}}]{SuperCDMS:2024yiv}%
  \BibitemOpen
  \bibfield  {author} {\bibinfo {author} {\bibfnamefont {M.~F.}\ \bibnamefont {Albakry}} \emph {et~al.} (\bibinfo {collaboration} {SuperCDMS}),\ }\href@noop {} {\bibinfo {title} {{Light Dark Matter Constraints from SuperCDMS HVeV Detectors Operated Underground with an Anticoincidence Event Selection}}} (\bibinfo {year} {2024}),\ \Eprint {https://arxiv.org/abs/2407.08085} {arXiv:2407.08085 [hep-ex]} \BibitemShut {NoStop}%
\bibitem [{\citenamefont {Hochberg}\ \emph {et~al.}(2021)\citenamefont {Hochberg}, \citenamefont {Kahn}, \citenamefont {Kurinsky}, \citenamefont {Lehmann}, \citenamefont {Yu},\ and\ \citenamefont {Berggren}}]{Hochberg:2021pkt}%
  \BibitemOpen
  \bibfield  {author} {\bibinfo {author} {\bibfnamefont {Y.}~\bibnamefont {Hochberg}}, \bibinfo {author} {\bibfnamefont {Y.}~\bibnamefont {Kahn}}, \bibinfo {author} {\bibfnamefont {N.}~\bibnamefont {Kurinsky}}, \bibinfo {author} {\bibfnamefont {B.~V.}\ \bibnamefont {Lehmann}}, \bibinfo {author} {\bibfnamefont {T.~C.}\ \bibnamefont {Yu}},\ and\ \bibinfo {author} {\bibfnamefont {K.~K.}\ \bibnamefont {Berggren}},\ }\bibfield  {title} {\bibinfo {title} {{Determining Dark-Matter\textendash{}Electron Scattering Rates from the Dielectric Function}},\ }\href {https://doi.org/10.1103/PhysRevLett.127.151802} {\bibfield  {journal} {\bibinfo  {journal} {Phys. Rev. Lett.}\ }\textbf {\bibinfo {volume} {127}},\ \bibinfo {pages} {151802} (\bibinfo {year} {2021})},\ \Eprint {https://arxiv.org/abs/2101.08263} {arXiv:2101.08263 [hep-ph]} \BibitemShut {NoStop}%
\bibitem [{\citenamefont {Knapen}\ \emph {et~al.}(2021)\citenamefont {Knapen}, \citenamefont {Kozaczuk},\ and\ \citenamefont {Lin}}]{Knapen:2021run}%
  \BibitemOpen
  \bibfield  {author} {\bibinfo {author} {\bibfnamefont {S.}~\bibnamefont {Knapen}}, \bibinfo {author} {\bibfnamefont {J.}~\bibnamefont {Kozaczuk}},\ and\ \bibinfo {author} {\bibfnamefont {T.}~\bibnamefont {Lin}},\ }\bibfield  {title} {\bibinfo {title} {{Dark matter-electron scattering in dielectrics}},\ }\href {https://doi.org/10.1103/PhysRevD.104.015031} {\bibfield  {journal} {\bibinfo  {journal} {Phys. Rev. D}\ }\textbf {\bibinfo {volume} {104}},\ \bibinfo {pages} {015031} (\bibinfo {year} {2021})},\ \Eprint {https://arxiv.org/abs/2101.08275} {arXiv:2101.08275 [hep-ph]} \BibitemShut {NoStop}%
\bibitem [{\citenamefont {Boyd}\ \emph {et~al.}(2023)\citenamefont {Boyd}, \citenamefont {Hochberg}, \citenamefont {Kahn}, \citenamefont {Kramer}, \citenamefont {Kurinsky}, \citenamefont {Lehmann},\ and\ \citenamefont {Yu}}]{Boyd:2022tcn}%
  \BibitemOpen
  \bibfield  {author} {\bibinfo {author} {\bibfnamefont {C.}~\bibnamefont {Boyd}}, \bibinfo {author} {\bibfnamefont {Y.}~\bibnamefont {Hochberg}}, \bibinfo {author} {\bibfnamefont {Y.}~\bibnamefont {Kahn}}, \bibinfo {author} {\bibfnamefont {E.~D.}\ \bibnamefont {Kramer}}, \bibinfo {author} {\bibfnamefont {N.}~\bibnamefont {Kurinsky}}, \bibinfo {author} {\bibfnamefont {B.~V.}\ \bibnamefont {Lehmann}},\ and\ \bibinfo {author} {\bibfnamefont {T.~C.}\ \bibnamefont {Yu}},\ }\bibfield  {title} {\bibinfo {title} {{Directional detection of dark matter with anisotropic response functions}},\ }\href {https://doi.org/10.1103/PhysRevD.108.015015} {\bibfield  {journal} {\bibinfo  {journal} {Phys. Rev. D}\ }\textbf {\bibinfo {volume} {108}},\ \bibinfo {pages} {015015} (\bibinfo {year} {2023})},\ \Eprint {https://arxiv.org/abs/2212.04505} {arXiv:2212.04505 [hep-ph]} \BibitemShut {NoStop}%
\bibitem [{\citenamefont {Jain}\ \emph {et~al.}(2013)\citenamefont {Jain} \emph {et~al.}}]{Jain:2013wst}%
  \BibitemOpen
  \bibfield  {author} {\bibinfo {author} {\bibfnamefont {A.}~\bibnamefont {Jain}} \emph {et~al.},\ }\bibfield  {title} {\bibinfo {title} {{Commentary: The Materials Project: A materials genome approach to accelerating materials innovation}},\ }\href {https://doi.org/10.1063/1.4812323} {\bibfield  {journal} {\bibinfo  {journal} {APL Mater.}\ }\textbf {\bibinfo {volume} {1}},\ \bibinfo {pages} {011002} (\bibinfo {year} {2013})}\BibitemShut {NoStop}%
\bibitem [{\citenamefont {Petousis}\ \emph {et~al.}(2017)\citenamefont {Petousis}, \citenamefont {Mrdjenovich}, \citenamefont {Ballouz}, \citenamefont {Liu}, \citenamefont {Winston}, \citenamefont {Chen}, \citenamefont {Graf}, \citenamefont {Schladt}, \citenamefont {Persson},\ and\ \citenamefont {Prinz}}]{Petousis:2017}%
  \BibitemOpen
  \bibfield  {author} {\bibinfo {author} {\bibfnamefont {I.}~\bibnamefont {Petousis}}, \bibinfo {author} {\bibfnamefont {D.}~\bibnamefont {Mrdjenovich}}, \bibinfo {author} {\bibfnamefont {E.}~\bibnamefont {Ballouz}}, \bibinfo {author} {\bibfnamefont {M.}~\bibnamefont {Liu}}, \bibinfo {author} {\bibfnamefont {D.}~\bibnamefont {Winston}}, \bibinfo {author} {\bibfnamefont {W.}~\bibnamefont {Chen}}, \bibinfo {author} {\bibfnamefont {T.}~\bibnamefont {Graf}}, \bibinfo {author} {\bibfnamefont {T.~D.}\ \bibnamefont {Schladt}}, \bibinfo {author} {\bibfnamefont {K.~A.}\ \bibnamefont {Persson}},\ and\ \bibinfo {author} {\bibfnamefont {F.~B.}\ \bibnamefont {Prinz}},\ }\bibfield  {title} {\bibinfo {title} {High-throughput screening of inorganic compounds for the discovery of novel dielectric and optical materials},\ }\href {https://doi.org/10.1038/sdata.2016.134} {\bibfield  {journal} {\bibinfo  {journal} {Scientific Data}\ }\textbf {\bibinfo {volume} {4}},\ \bibinfo {pages} {160134} (\bibinfo {year} {2017})}\BibitemShut
  {NoStop}%
\bibitem [{\citenamefont {{Yang}}\ \emph {et~al.}()\citenamefont {{Yang}}, \citenamefont {{Horton}}, \citenamefont {{Munro}},\ and\ \citenamefont {{Persson}}}]{Ruoxi:2022}%
  \BibitemOpen
  \bibfield  {author} {\bibinfo {author} {\bibfnamefont {R.~X.}\ \bibnamefont {{Yang}}}, \bibinfo {author} {\bibfnamefont {M.~K.}\ \bibnamefont {{Horton}}}, \bibinfo {author} {\bibfnamefont {J.}~\bibnamefont {{Munro}}},\ and\ \bibinfo {author} {\bibfnamefont {K.~A.}\ \bibnamefont {{Persson}}},\ }\bibfield  {title} {\bibinfo {title} {{High-throughput optical absorption spectra for inorganic semiconductors}},\ }\href@noop {} {\ }\Eprint {https://arxiv.org/abs/arXiv:2209.02918} {arXiv:2209.02918 [cond-mat.mtrl-sci]} \BibitemShut {NoStop}%
\bibitem [{\citenamefont {Inzani}\ \emph {et~al.}(2021)\citenamefont {Inzani}, \citenamefont {Faghaninia},\ and\ \citenamefont {Griffin}}]{Inzani:2020szg}%
  \BibitemOpen
  \bibfield  {author} {\bibinfo {author} {\bibfnamefont {K.}~\bibnamefont {Inzani}}, \bibinfo {author} {\bibfnamefont {A.}~\bibnamefont {Faghaninia}},\ and\ \bibinfo {author} {\bibfnamefont {S.~M.}\ \bibnamefont {Griffin}},\ }\bibfield  {title} {\bibinfo {title} {{Prediction of Tunable Spin-Orbit Gapped Materials for Dark Matter Detection}},\ }\href {https://doi.org/10.1103/PhysRevResearch.3.013069} {\bibfield  {journal} {\bibinfo  {journal} {Phys. Rev. Res.}\ }\textbf {\bibinfo {volume} {3}},\ \bibinfo {pages} {013069} (\bibinfo {year} {2021})},\ \Eprint {https://arxiv.org/abs/2008.05062} {arXiv:2008.05062 [cond-mat.mtrl-sci]} \BibitemShut {NoStop}%
\bibitem [{\citenamefont {Geilhufe}\ \emph {et~al.}(2018)\citenamefont {Geilhufe}, \citenamefont {Olsthoorn}, \citenamefont {Ferella}, \citenamefont {Koski}, \citenamefont {Kahlhoefer}, \citenamefont {Conrad},\ and\ \citenamefont {Balatsky}}]{Geilhufe:2018gry}%
  \BibitemOpen
  \bibfield  {author} {\bibinfo {author} {\bibfnamefont {R.~M.}\ \bibnamefont {Geilhufe}}, \bibinfo {author} {\bibfnamefont {B.}~\bibnamefont {Olsthoorn}}, \bibinfo {author} {\bibfnamefont {A.}~\bibnamefont {Ferella}}, \bibinfo {author} {\bibfnamefont {T.}~\bibnamefont {Koski}}, \bibinfo {author} {\bibfnamefont {F.}~\bibnamefont {Kahlhoefer}}, \bibinfo {author} {\bibfnamefont {J.}~\bibnamefont {Conrad}},\ and\ \bibinfo {author} {\bibfnamefont {A.~V.}\ \bibnamefont {Balatsky}},\ }\bibfield  {title} {\bibinfo {title} {{Materials Informatics for Dark Matter Detection}},\ }\href {https://doi.org/10.1002/pssr.201800293} {\bibfield  {journal} {\bibinfo  {journal} {Phys. Status Solidi RRL}\ }\textbf {\bibinfo {volume} {12}},\ \bibinfo {pages} {1800293} (\bibinfo {year} {2018})},\ \Eprint {https://arxiv.org/abs/1806.06040} {arXiv:1806.06040 [cond-mat.mtrl-sci]} \BibitemShut {NoStop}%
\bibitem [{\citenamefont {Hochberg}\ \emph {et~al.}(2018)\citenamefont {Hochberg}, \citenamefont {Kahn}, \citenamefont {Lisanti}, \citenamefont {Zurek}, \citenamefont {Grushin}, \citenamefont {Ilan}, \citenamefont {Griffin}, \citenamefont {Liu}, \citenamefont {Weber},\ and\ \citenamefont {Neaton}}]{Hochberg:2017wce}%
  \BibitemOpen
  \bibfield  {author} {\bibinfo {author} {\bibfnamefont {Y.}~\bibnamefont {Hochberg}}, \bibinfo {author} {\bibfnamefont {Y.}~\bibnamefont {Kahn}}, \bibinfo {author} {\bibfnamefont {M.}~\bibnamefont {Lisanti}}, \bibinfo {author} {\bibfnamefont {K.~M.}\ \bibnamefont {Zurek}}, \bibinfo {author} {\bibfnamefont {A.~G.}\ \bibnamefont {Grushin}}, \bibinfo {author} {\bibfnamefont {R.}~\bibnamefont {Ilan}}, \bibinfo {author} {\bibfnamefont {S.~M.}\ \bibnamefont {Griffin}}, \bibinfo {author} {\bibfnamefont {Z.-F.}\ \bibnamefont {Liu}}, \bibinfo {author} {\bibfnamefont {S.~F.}\ \bibnamefont {Weber}},\ and\ \bibinfo {author} {\bibfnamefont {J.~B.}\ \bibnamefont {Neaton}},\ }\bibfield  {title} {\bibinfo {title} {{Detection of sub-MeV Dark Matter with Three-Dimensional Dirac Materials}},\ }\href {https://doi.org/10.1103/PhysRevD.97.015004} {\bibfield  {journal} {\bibinfo  {journal} {Phys. Rev. D}\ }\textbf {\bibinfo {volume} {97}},\ \bibinfo {pages} {015004} (\bibinfo {year} {2018})},\ \Eprint
  {https://arxiv.org/abs/1708.08929} {arXiv:1708.08929 [hep-ph]} \BibitemShut {NoStop}%
\bibitem [{\citenamefont {Budnik}\ \emph {et~al.}(2018)\citenamefont {Budnik}, \citenamefont {Chesnovsky}, \citenamefont {Slone},\ and\ \citenamefont {Volansky}}]{Budnik:2017sbu}%
  \BibitemOpen
  \bibfield  {author} {\bibinfo {author} {\bibfnamefont {R.}~\bibnamefont {Budnik}}, \bibinfo {author} {\bibfnamefont {O.}~\bibnamefont {Chesnovsky}}, \bibinfo {author} {\bibfnamefont {O.}~\bibnamefont {Slone}},\ and\ \bibinfo {author} {\bibfnamefont {T.}~\bibnamefont {Volansky}},\ }\bibfield  {title} {\bibinfo {title} {{Direct Detection of Light Dark Matter and Solar Neutrinos via Color Center Production in Crystals}},\ }\href {https://doi.org/10.1016/j.physletb.2018.04.063} {\bibfield  {journal} {\bibinfo  {journal} {Phys. Lett. B}\ }\textbf {\bibinfo {volume} {782}},\ \bibinfo {pages} {242} (\bibinfo {year} {2018})},\ \Eprint {https://arxiv.org/abs/1705.03016} {arXiv:1705.03016 [hep-ph]} \BibitemShut {NoStop}%
\bibitem [{\citenamefont {Kadribasic}\ \emph {et~al.}(2018)\citenamefont {Kadribasic}, \citenamefont {Mirabolfathi}, \citenamefont {Nordlund}, \citenamefont {Sand}, \citenamefont {Holmstr\"om},\ and\ \citenamefont {Djurabekova}}]{Kadribasic:2017obi}%
  \BibitemOpen
  \bibfield  {author} {\bibinfo {author} {\bibfnamefont {F.}~\bibnamefont {Kadribasic}}, \bibinfo {author} {\bibfnamefont {N.}~\bibnamefont {Mirabolfathi}}, \bibinfo {author} {\bibfnamefont {K.}~\bibnamefont {Nordlund}}, \bibinfo {author} {\bibfnamefont {A.~E.}\ \bibnamefont {Sand}}, \bibinfo {author} {\bibfnamefont {E.}~\bibnamefont {Holmstr\"om}},\ and\ \bibinfo {author} {\bibfnamefont {F.}~\bibnamefont {Djurabekova}},\ }\bibfield  {title} {\bibinfo {title} {{Directional Sensitivity In Light-Mass Dark Matter Searches With Single-Electron Resolution Ionization Detectors}},\ }\href {https://doi.org/10.1103/PhysRevLett.120.111301} {\bibfield  {journal} {\bibinfo  {journal} {Phys. Rev. Lett.}\ }\textbf {\bibinfo {volume} {120}},\ \bibinfo {pages} {111301} (\bibinfo {year} {2018})},\ \Eprint {https://arxiv.org/abs/1703.05371} {arXiv:1703.05371 [physics.ins-det]} \BibitemShut {NoStop}%
\bibitem [{\citenamefont {Griffin}\ \emph {et~al.}(2018)\citenamefont {Griffin}, \citenamefont {Knapen}, \citenamefont {Lin},\ and\ \citenamefont {Zurek}}]{Griffin:2018bjn}%
  \BibitemOpen
  \bibfield  {author} {\bibinfo {author} {\bibfnamefont {S.}~\bibnamefont {Griffin}}, \bibinfo {author} {\bibfnamefont {S.}~\bibnamefont {Knapen}}, \bibinfo {author} {\bibfnamefont {T.}~\bibnamefont {Lin}},\ and\ \bibinfo {author} {\bibfnamefont {K.~M.}\ \bibnamefont {Zurek}},\ }\bibfield  {title} {\bibinfo {title} {{Directional Detection of Light Dark Matter with Polar Materials}},\ }\href {https://doi.org/10.1103/PhysRevD.98.115034} {\bibfield  {journal} {\bibinfo  {journal} {Phys. Rev. D}\ }\textbf {\bibinfo {volume} {98}},\ \bibinfo {pages} {115034} (\bibinfo {year} {2018})},\ \Eprint {https://arxiv.org/abs/1807.10291} {arXiv:1807.10291 [hep-ph]} \BibitemShut {NoStop}%
\bibitem [{\citenamefont {Heikinheimo}\ \emph {et~al.}(2019)\citenamefont {Heikinheimo}, \citenamefont {Nordlund}, \citenamefont {Tuominen},\ and\ \citenamefont {Mirabolfathi}}]{Heikinheimo:2019lwg}%
  \BibitemOpen
  \bibfield  {author} {\bibinfo {author} {\bibfnamefont {M.}~\bibnamefont {Heikinheimo}}, \bibinfo {author} {\bibfnamefont {K.}~\bibnamefont {Nordlund}}, \bibinfo {author} {\bibfnamefont {K.}~\bibnamefont {Tuominen}},\ and\ \bibinfo {author} {\bibfnamefont {N.}~\bibnamefont {Mirabolfathi}},\ }\bibfield  {title} {\bibinfo {title} {{Velocity Dependent Dark Matter Interactions in Single-Electron Resolution Semiconductor Detectors with Directional Sensitivity}},\ }\href {https://doi.org/10.1103/PhysRevD.99.103018} {\bibfield  {journal} {\bibinfo  {journal} {Phys. Rev. D}\ }\textbf {\bibinfo {volume} {99}},\ \bibinfo {pages} {103018} (\bibinfo {year} {2019})},\ \Eprint {https://arxiv.org/abs/1903.08654} {arXiv:1903.08654 [hep-ph]} \BibitemShut {NoStop}%
\bibitem [{\citenamefont {Coskuner}\ \emph {et~al.}(2021)\citenamefont {Coskuner}, \citenamefont {Mitridate}, \citenamefont {Olivares},\ and\ \citenamefont {Zurek}}]{Coskuner:2019odd}%
  \BibitemOpen
  \bibfield  {author} {\bibinfo {author} {\bibfnamefont {A.}~\bibnamefont {Coskuner}}, \bibinfo {author} {\bibfnamefont {A.}~\bibnamefont {Mitridate}}, \bibinfo {author} {\bibfnamefont {A.}~\bibnamefont {Olivares}},\ and\ \bibinfo {author} {\bibfnamefont {K.~M.}\ \bibnamefont {Zurek}},\ }\bibfield  {title} {\bibinfo {title} {{Directional Dark Matter Detection in Anisotropic Dirac Materials}},\ }\href {https://doi.org/10.1103/PhysRevD.103.016006} {\bibfield  {journal} {\bibinfo  {journal} {Phys. Rev. D}\ }\textbf {\bibinfo {volume} {103}},\ \bibinfo {pages} {016006} (\bibinfo {year} {2021})},\ \Eprint {https://arxiv.org/abs/1909.09170} {arXiv:1909.09170 [hep-ph]} \BibitemShut {NoStop}%
\bibitem [{\citenamefont {Geilhufe}\ \emph {et~al.}(2020)\citenamefont {Geilhufe}, \citenamefont {Kahlhoefer},\ and\ \citenamefont {Winkler}}]{Geilhufe:2019ndy}%
  \BibitemOpen
  \bibfield  {author} {\bibinfo {author} {\bibfnamefont {R.~M.}\ \bibnamefont {Geilhufe}}, \bibinfo {author} {\bibfnamefont {F.}~\bibnamefont {Kahlhoefer}},\ and\ \bibinfo {author} {\bibfnamefont {M.~W.}\ \bibnamefont {Winkler}},\ }\bibfield  {title} {\bibinfo {title} {{Dirac Materials for Sub-MeV Dark Matter Detection: New Targets and Improved Formalism}},\ }\href {https://doi.org/10.1103/PhysRevD.101.055005} {\bibfield  {journal} {\bibinfo  {journal} {Phys. Rev. D}\ }\textbf {\bibinfo {volume} {101}},\ \bibinfo {pages} {055005} (\bibinfo {year} {2020})},\ \Eprint {https://arxiv.org/abs/1910.02091} {arXiv:1910.02091 [hep-ph]} \BibitemShut {NoStop}%
\bibitem [{\citenamefont {Coskuner}\ \emph {et~al.}(2022)\citenamefont {Coskuner}, \citenamefont {Trickle}, \citenamefont {Zhang},\ and\ \citenamefont {Zurek}}]{Coskuner:2021qxo}%
  \BibitemOpen
  \bibfield  {author} {\bibinfo {author} {\bibfnamefont {A.}~\bibnamefont {Coskuner}}, \bibinfo {author} {\bibfnamefont {T.}~\bibnamefont {Trickle}}, \bibinfo {author} {\bibfnamefont {Z.}~\bibnamefont {Zhang}},\ and\ \bibinfo {author} {\bibfnamefont {K.~M.}\ \bibnamefont {Zurek}},\ }\bibfield  {title} {\bibinfo {title} {{Directional detectability of dark matter with single phonon excitations: Target comparison}},\ }\href {https://doi.org/10.1103/PhysRevD.105.015010} {\bibfield  {journal} {\bibinfo  {journal} {Phys. Rev. D}\ }\textbf {\bibinfo {volume} {105}},\ \bibinfo {pages} {015010} (\bibinfo {year} {2022})},\ \Eprint {https://arxiv.org/abs/2102.09567} {arXiv:2102.09567 [hep-ph]} \BibitemShut {NoStop}%
\bibitem [{\citenamefont {Sassi}\ \emph {et~al.}(2021)\citenamefont {Sassi}, \citenamefont {Dinmohammadi}, \citenamefont {Heikinheimo}, \citenamefont {Mirabolfathi}, \citenamefont {Nordlund}, \citenamefont {Safari},\ and\ \citenamefont {Tuominen}}]{Sassi:2021umf}%
  \BibitemOpen
  \bibfield  {author} {\bibinfo {author} {\bibfnamefont {S.}~\bibnamefont {Sassi}}, \bibinfo {author} {\bibfnamefont {A.}~\bibnamefont {Dinmohammadi}}, \bibinfo {author} {\bibfnamefont {M.}~\bibnamefont {Heikinheimo}}, \bibinfo {author} {\bibfnamefont {N.}~\bibnamefont {Mirabolfathi}}, \bibinfo {author} {\bibfnamefont {K.}~\bibnamefont {Nordlund}}, \bibinfo {author} {\bibfnamefont {H.}~\bibnamefont {Safari}},\ and\ \bibinfo {author} {\bibfnamefont {K.}~\bibnamefont {Tuominen}},\ }\bibfield  {title} {\bibinfo {title} {{Solar neutrinos and dark matter detection with diurnal modulation}},\ }\href {https://doi.org/10.1103/PhysRevD.104.063037} {\bibfield  {journal} {\bibinfo  {journal} {Phys. Rev. D}\ }\textbf {\bibinfo {volume} {104}},\ \bibinfo {pages} {063037} (\bibinfo {year} {2021})},\ \Eprint {https://arxiv.org/abs/2103.08511} {arXiv:2103.08511 [hep-ph]} \BibitemShut {NoStop}%
\bibitem [{\citenamefont {Hochberg}\ \emph {et~al.}(2023)\citenamefont {Hochberg}, \citenamefont {Kramer}, \citenamefont {Kurinsky},\ and\ \citenamefont {Lehmann}}]{Hochberg:2021ymx}%
  \BibitemOpen
  \bibfield  {author} {\bibinfo {author} {\bibfnamefont {Y.}~\bibnamefont {Hochberg}}, \bibinfo {author} {\bibfnamefont {E.~D.}\ \bibnamefont {Kramer}}, \bibinfo {author} {\bibfnamefont {N.}~\bibnamefont {Kurinsky}},\ and\ \bibinfo {author} {\bibfnamefont {B.~V.}\ \bibnamefont {Lehmann}},\ }\bibfield  {title} {\bibinfo {title} {{Directional detection of light dark matter in superconductors}},\ }\href {https://doi.org/10.1103/PhysRevD.107.076015} {\bibfield  {journal} {\bibinfo  {journal} {Phys. Rev. D}\ }\textbf {\bibinfo {volume} {107}},\ \bibinfo {pages} {076015} (\bibinfo {year} {2023})},\ \Eprint {https://arxiv.org/abs/2109.04473} {arXiv:2109.04473 [hep-ph]} \BibitemShut {NoStop}%
\bibitem [{\citenamefont {Blanco}\ \emph {et~al.}(2022)\citenamefont {Blanco}, \citenamefont {Harris}, \citenamefont {Kahn}, \citenamefont {Lillard},\ and\ \citenamefont {P\'erez-R\'\i{}os}}]{Blanco:2022pkt}%
  \BibitemOpen
  \bibfield  {author} {\bibinfo {author} {\bibfnamefont {C.}~\bibnamefont {Blanco}}, \bibinfo {author} {\bibfnamefont {I.}~\bibnamefont {Harris}}, \bibinfo {author} {\bibfnamefont {Y.}~\bibnamefont {Kahn}}, \bibinfo {author} {\bibfnamefont {B.}~\bibnamefont {Lillard}},\ and\ \bibinfo {author} {\bibfnamefont {J.}~\bibnamefont {P\'erez-R\'\i{}os}},\ }\bibfield  {title} {\bibinfo {title} {{Molecular Migdal effect}},\ }\href {https://doi.org/10.1103/PhysRevD.106.115015} {\bibfield  {journal} {\bibinfo  {journal} {Phys. Rev. D}\ }\textbf {\bibinfo {volume} {106}},\ \bibinfo {pages} {115015} (\bibinfo {year} {2022})},\ \Eprint {https://arxiv.org/abs/2208.09002} {arXiv:2208.09002 [hep-ph]} \BibitemShut {NoStop}%
\bibitem [{\citenamefont {Spergel}(1988)}]{Spergel:1987kx}%
  \BibitemOpen
  \bibfield  {author} {\bibinfo {author} {\bibfnamefont {D.~N.}\ \bibnamefont {Spergel}},\ }\bibfield  {title} {\bibinfo {title} {{The Motion of the Earth and the Detection of Wimps}},\ }\href {https://doi.org/10.1103/PhysRevD.37.1353} {\bibfield  {journal} {\bibinfo  {journal} {Phys. Rev. D}\ }\textbf {\bibinfo {volume} {37}},\ \bibinfo {pages} {1353} (\bibinfo {year} {1988})}\BibitemShut {NoStop}%
\bibitem [{\citenamefont {Mayet}\ \emph {et~al.}(2016)\citenamefont {Mayet} \emph {et~al.}}]{Mayet:2016zxu}%
  \BibitemOpen
  \bibfield  {author} {\bibinfo {author} {\bibfnamefont {F.}~\bibnamefont {Mayet}} \emph {et~al.},\ }\bibfield  {title} {\bibinfo {title} {{A review of the discovery reach of directional Dark Matter detection}},\ }\href {https://doi.org/10.1016/j.physrep.2016.02.007} {\bibfield  {journal} {\bibinfo  {journal} {Phys. Rept.}\ }\textbf {\bibinfo {volume} {627}},\ \bibinfo {pages} {1} (\bibinfo {year} {2016})},\ \Eprint {https://arxiv.org/abs/1602.03781} {arXiv:1602.03781 [astro-ph.CO]} \BibitemShut {NoStop}%
\bibitem [{\citenamefont {Lewin}\ and\ \citenamefont {Smith}(1996)}]{Lewin:1995rx}%
  \BibitemOpen
  \bibfield  {author} {\bibinfo {author} {\bibfnamefont {J.~D.}\ \bibnamefont {Lewin}}\ and\ \bibinfo {author} {\bibfnamefont {P.~F.}\ \bibnamefont {Smith}},\ }\bibfield  {title} {\bibinfo {title} {{Review of mathematics, numerical factors, and corrections for dark matter experiments based on elastic nuclear recoil}},\ }\href {https://doi.org/10.1016/S0927-6505(96)00047-3} {\bibfield  {journal} {\bibinfo  {journal} {Astropart. Phys.}\ }\textbf {\bibinfo {volume} {6}},\ \bibinfo {pages} {87} (\bibinfo {year} {1996})}\BibitemShut {NoStop}%
\bibitem [{\citenamefont {Knapen}\ \emph {et~al.}(2022)\citenamefont {Knapen}, \citenamefont {Kozaczuk},\ and\ \citenamefont {Lin}}]{Knapen:2021bwg}%
  \BibitemOpen
  \bibfield  {author} {\bibinfo {author} {\bibfnamefont {S.}~\bibnamefont {Knapen}}, \bibinfo {author} {\bibfnamefont {J.}~\bibnamefont {Kozaczuk}},\ and\ \bibinfo {author} {\bibfnamefont {T.}~\bibnamefont {Lin}},\ }\bibfield  {title} {\bibinfo {title} {{python package for dark matter scattering in dielectric targets}},\ }\href {https://doi.org/10.1103/PhysRevD.105.015014} {\bibfield  {journal} {\bibinfo  {journal} {Phys. Rev. D}\ }\textbf {\bibinfo {volume} {105}},\ \bibinfo {pages} {015014} (\bibinfo {year} {2022})},\ \Eprint {https://arxiv.org/abs/2104.12786} {arXiv:2104.12786 [hep-ph]} \BibitemShut {NoStop}%
\bibitem [{\citenamefont {Lehmann}\ \emph {et~al.}()\citenamefont {Lehmann} \emph {et~al.}}]{to-appear}%
  \BibitemOpen
  \bibfield  {author} {\bibinfo {author} {\bibfnamefont {B.~V.}\ \bibnamefont {Lehmann}} \emph {et~al.},\ }\href@noop {} {\bibinfo  {journal} {to appear}\ }\BibitemShut {NoStop}%
\bibitem [{\citenamefont {Griffin}\ \emph {et~al.}()\citenamefont {Griffin}, \citenamefont {Hochberg}, \citenamefont {Lehmann}, \citenamefont {Ovadia}, \citenamefont {Suter},\ and\ \citenamefont {Yang}}]{full_repository}%
  \BibitemOpen
\bibfield  {journal} {  }\bibfield  {author} {\bibinfo {author} {\bibfnamefont {S.~M.}\ \bibnamefont {Griffin}}, \bibinfo {author} {\bibfnamefont {Y.}~\bibnamefont {Hochberg}}, \bibinfo {author} {\bibfnamefont {B.~V.}\ \bibnamefont {Lehmann}}, \bibinfo {author} {\bibfnamefont {R.}~\bibnamefont {Ovadia}}, \bibinfo {author} {\bibfnamefont {B.~A.}\ \bibnamefont {Suter}},\ and\ \bibinfo {author} {\bibfnamefont {R.~X.}\ \bibnamefont {Yang}},\ }\bibfield  {title} {\bibinfo {title} {{Dark matter sensitivity projections from the Materials Project dataset}},\ }\href {https://doi.org/10.5281/zenodo.13346342} {10.5281/zenodo.13346342}\BibitemShut {NoStop}%
\bibitem [{\citenamefont {Dressel}\ and\ \citenamefont {Grüner}(2002)}]{Dressel_Gruner:2002}%
  \BibitemOpen
  \bibfield  {author} {\bibinfo {author} {\bibfnamefont {M.}~\bibnamefont {Dressel}}\ and\ \bibinfo {author} {\bibfnamefont {G.}~\bibnamefont {Grüner}},\ }\href@noop {} {\emph {\bibinfo {title} {{Electrodynamics of Solids: Optical Properties of Electrons in Matter}}}}\ (\bibinfo  {publisher} {{Cambridge University Press}},\ \bibinfo {year} {2002})\BibitemShut {NoStop}%
\bibitem [{\citenamefont {Abril}\ \emph {et~al.}(1998)\citenamefont {Abril}, \citenamefont {Garcia-Molina}, \citenamefont {Denton}, \citenamefont {P\'erez-P\'erez},\ and\ \citenamefont {Arista}}]{Abril:1998rea}%
  \BibitemOpen
  \bibfield  {author} {\bibinfo {author} {\bibfnamefont {I.}~\bibnamefont {Abril}}, \bibinfo {author} {\bibfnamefont {R.}~\bibnamefont {Garcia-Molina}}, \bibinfo {author} {\bibfnamefont {C.~D.}\ \bibnamefont {Denton}}, \bibinfo {author} {\bibfnamefont {F.~J.}\ \bibnamefont {P\'erez-P\'erez}},\ and\ \bibinfo {author} {\bibfnamefont {N.~R.}\ \bibnamefont {Arista}},\ }\bibfield  {title} {\bibinfo {title} {Dielectric description of wakes and stopping powers in solids},\ }\href {https://doi.org/10.1103/PhysRevA.58.357} {\bibfield  {journal} {\bibinfo  {journal} {Phys. Rev. A}\ }\textbf {\bibinfo {volume} {58}},\ \bibinfo {pages} {357} (\bibinfo {year} {1998})}\BibitemShut {NoStop}%
\bibitem [{\citenamefont {{Vos}}\ and\ \citenamefont {{Grande}}(2019)}]{2019JPCS..124..242V}%
  \BibitemOpen
  \bibfield  {author} {\bibinfo {author} {\bibfnamefont {M.}~\bibnamefont {{Vos}}}\ and\ \bibinfo {author} {\bibfnamefont {P.~L.}\ \bibnamefont {{Grande}}},\ }\bibfield  {title} {\bibinfo {title} {{Modelling the contribution of semi-core electrons to the dielectric function}},\ }\href {https://doi.org/10.1016/j.jpcs.2018.09.020} {\bibfield  {journal} {\bibinfo  {journal} {Journal of Physics and Chemistry of Solids}\ }\textbf {\bibinfo {volume} {124}},\ \bibinfo {pages} {242} (\bibinfo {year} {2019})}\BibitemShut {NoStop}%
\bibitem [{\citenamefont {Hochberg}\ \emph {et~al.}(2025)\citenamefont {Hochberg}, \citenamefont {Novko}, \citenamefont {Ovadia},\ and\ \citenamefont {Politano}}]{Dino:future}%
  \BibitemOpen
  \bibfield  {author} {\bibinfo {author} {\bibfnamefont {Y.}~\bibnamefont {Hochberg}}, \bibinfo {author} {\bibfnamefont {D.}~\bibnamefont {Novko}}, \bibinfo {author} {\bibfnamefont {R.}~\bibnamefont {Ovadia}},\ and\ \bibinfo {author} {\bibfnamefont {A.}~\bibnamefont {Politano}},\ }\href@noop {} {\bibfield  {journal} {\bibinfo  {journal} {to appear}\ } (\bibinfo {year} {2025})}\BibitemShut {NoStop}%
\bibitem [{\citenamefont {{Luu}}\ and\ \citenamefont {{Vaqueiro}}(2015)}]{2015JSSCh.226..219L}%
  \BibitemOpen
  \bibfield  {author} {\bibinfo {author} {\bibfnamefont {S.~D.~N.}\ \bibnamefont {{Luu}}}\ and\ \bibinfo {author} {\bibfnamefont {P.}~\bibnamefont {{Vaqueiro}}},\ }\bibfield  {title} {\bibinfo {title} {{Synthesis, characterisation and thermoelectric properties of the oxytelluride Bi$_{2}$O$_{2}$Te}},\ }\href {https://doi.org/10.1016/j.jssc.2015.02.026} {\bibfield  {journal} {\bibinfo  {journal} {Journal of Solid State Chemistry France}\ }\textbf {\bibinfo {volume} {226}},\ \bibinfo {pages} {219} (\bibinfo {year} {2015})}\BibitemShut {NoStop}%
\bibitem [{\citenamefont {{Sands}}\ \emph {et~al.}(1963)\citenamefont {{Sands}}, \citenamefont {{Woods}},\ and\ \citenamefont {{Ramsey}}}]{1963AcCry..16..316S}%
  \BibitemOpen
  \bibfield  {author} {\bibinfo {author} {\bibfnamefont {D.~E.}\ \bibnamefont {{Sands}}}, \bibinfo {author} {\bibfnamefont {D.~H.}\ \bibnamefont {{Woods}}},\ and\ \bibinfo {author} {\bibfnamefont {W.~J.}\ \bibnamefont {{Ramsey}}},\ }\bibfield  {title} {\bibinfo {title} {{The crystal structure of {\ensuremath{\beta}}-\ce{K3Bi}}},\ }\href {https://doi.org/10.1107/S0365110X63000839} {\bibfield  {journal} {\bibinfo  {journal} {Acta Crystallographica}\ }\textbf {\bibinfo {volume} {16}},\ \bibinfo {pages} {316} (\bibinfo {year} {1963})}\BibitemShut {NoStop}%
\bibitem [{\citenamefont {{Nixon}}\ \emph {et~al.}(1966)\citenamefont {{Nixon}}, \citenamefont {{Parry}},\ and\ \citenamefont {{Ubbelohde}}}]{1966RSPSA.291..324N}%
  \BibitemOpen
  \bibfield  {author} {\bibinfo {author} {\bibfnamefont {D.~E.}\ \bibnamefont {{Nixon}}}, \bibinfo {author} {\bibfnamefont {G.~S.}\ \bibnamefont {{Parry}}},\ and\ \bibinfo {author} {\bibfnamefont {A.~R.}\ \bibnamefont {{Ubbelohde}}},\ }\bibfield  {title} {\bibinfo {title} {{Order-Disorder Transformations in Graphite Nitrates}},\ }\href {https://doi.org/10.1098/rspa.1966.0098} {\bibfield  {journal} {\bibinfo  {journal} {Proceedings of the Royal Society of London Series A}\ }\textbf {\bibinfo {volume} {291}},\ \bibinfo {pages} {324} (\bibinfo {year} {1966})}\BibitemShut {NoStop}%
\bibitem [{\citenamefont {Kr\"amer}\ \emph {et~al.}(1989)\citenamefont {Kr\"amer}, \citenamefont {Schleid}, \citenamefont {Schulze}, \citenamefont {Urland},\ and\ \citenamefont {Meyer}}]{Kramer:1989}%
  \BibitemOpen
  \bibfield  {author} {\bibinfo {author} {\bibfnamefont {K.}~\bibnamefont {Kr\"amer}}, \bibinfo {author} {\bibfnamefont {T.}~\bibnamefont {Schleid}}, \bibinfo {author} {\bibfnamefont {M.}~\bibnamefont {Schulze}}, \bibinfo {author} {\bibfnamefont {W.}~\bibnamefont {Urland}},\ and\ \bibinfo {author} {\bibfnamefont {G.}~\bibnamefont {Meyer}},\ }\bibfield  {title} {\bibinfo {title} {{Three Bromides of Lanthanum: \ce{LaBr2}, \ce{La2Br5}, and \ce{LaBr3}}},\ }\href {https://doi.org/10.1002/zaac.19895750109} {\bibfield  {journal} {\bibinfo  {journal} {Zeitschrift für anorganische und allgemeine Chemie}\ }\textbf {\bibinfo {volume} {575}},\ \bibinfo {pages} {61} (\bibinfo {year} {1989})}\BibitemShut {NoStop}%
\bibitem [{\citenamefont {Widera}\ and\ \citenamefont {Sch\"afer}(1980)}]{Widera:1980}%
  \BibitemOpen
  \bibfield  {author} {\bibinfo {author} {\bibfnamefont {A.}~\bibnamefont {Widera}}\ and\ \bibinfo {author} {\bibfnamefont {H.}~\bibnamefont {Sch\"afer}},\ }\bibfield  {title} {\bibinfo {title} {{\"Ubergangsformen zwischen zintlphasen und echten salzen: Die verbindungen \ce{A3BO} (mit $\ce{A}=\ce{Ca},\ce{Sr},\ce{Ba}$ und $\ce{B} = \ce{Sn}, \ce{Pb}$)}},\ }\href {https://doi.org/https://doi.org/10.1016/0025-5408(80)90200-7} {\bibfield  {journal} {\bibinfo  {journal} {Materials Research Bulletin}\ }\textbf {\bibinfo {volume} {15}},\ \bibinfo {pages} {1805} (\bibinfo {year} {1980})}\BibitemShut {NoStop}%
\bibitem [{\citenamefont {Down}\ \emph {et~al.}(1978)\citenamefont {Down}, \citenamefont {Haley}, \citenamefont {Hubberstey}, \citenamefont {Pulham},\ and\ \citenamefont {Thunder}}]{Down:1978}%
  \BibitemOpen
  \bibfield  {author} {\bibinfo {author} {\bibfnamefont {M.~G.}\ \bibnamefont {Down}}, \bibinfo {author} {\bibfnamefont {M.~J.}\ \bibnamefont {Haley}}, \bibinfo {author} {\bibfnamefont {P.}~\bibnamefont {Hubberstey}}, \bibinfo {author} {\bibfnamefont {R.~J.}\ \bibnamefont {Pulham}},\ and\ \bibinfo {author} {\bibfnamefont {A.~E.}\ \bibnamefont {Thunder}},\ }\bibfield  {title} {\bibinfo {title} {{Solutions of lithium salts in liquid lithium: preparation and X-ray crystal structure of the dilithium salt of carbodi-imide (cyanamide)}},\ }\href {https://doi.org/10.1039/DT9780001407} {\bibfield  {journal} {\bibinfo  {journal} {{J. Chem. Soc., Dalton Trans.}}\ }\textbf {\bibinfo {volume} {10}},\ \bibinfo {pages} {1407} (\bibinfo {year} {1978})}\BibitemShut {NoStop}%
\bibitem [{\citenamefont {Johrendt}\ \emph {et~al.}(1996)\citenamefont {Johrendt}, \citenamefont {Miericke},\ and\ \citenamefont {Mewis}}]{Johrendt:1996}%
  \BibitemOpen
  \bibfield  {author} {\bibinfo {author} {\bibfnamefont {D.}~\bibnamefont {Johrendt}}, \bibinfo {author} {\bibfnamefont {R.}~\bibnamefont {Miericke}},\ and\ \bibinfo {author} {\bibfnamefont {A.}~\bibnamefont {Mewis}},\ }\bibfield  {title} {\bibinfo {title} {{Neue Pnictide mit modifizierten \ce{AIB2}-Strukturen / New Pnictides with Modified \ce{AIB2}-Type Structures}},\ }\href {https://doi.org/doi:10.1515/znb-1996-0625} {\bibfield  {journal} {\bibinfo  {journal} {Zeitschrift für Naturforschung B}\ }\textbf {\bibinfo {volume} {51}},\ \bibinfo {pages} {905} (\bibinfo {year} {1996})}\BibitemShut {NoStop}%
\bibitem [{\citenamefont {Bijvoet}\ \emph {et~al.}(1926)\citenamefont {Bijvoet}, \citenamefont {Claassen},\ and\ \citenamefont {Karssen}}]{Bijvoet:1926}%
  \BibitemOpen
  \bibfield  {author} {\bibinfo {author} {\bibfnamefont {J.}~\bibnamefont {Bijvoet}}, \bibinfo {author} {\bibfnamefont {A.}~\bibnamefont {Claassen}},\ and\ \bibinfo {author} {\bibfnamefont {A.}~\bibnamefont {Karssen}},\ }\bibfield  {title} {\bibinfo {title} {The scattering power of lithium and oxygen, determined from the diffraction-intensities of powdered lithiumoxide},\ }in\ \href@noop {} {\emph {\bibinfo {booktitle} {Proceedings of the Koninklijke Nederlandse Academie van Wetenschappen}}},\ Vol.~\bibinfo {volume} {29}\ (\bibinfo {year} {1926})\ pp.\ \bibinfo {pages} {1286--1292}\BibitemShut {NoStop}%
\bibitem [{\citenamefont {Rabenau}\ and\ \citenamefont {Schulz}(1976)}]{Rabenau:1976}%
  \BibitemOpen
  \bibfield  {author} {\bibinfo {author} {\bibfnamefont {A.}~\bibnamefont {Rabenau}}\ and\ \bibinfo {author} {\bibfnamefont {H.}~\bibnamefont {Schulz}},\ }\bibfield  {title} {\bibinfo {title} {Re-evaluation of the lithium nitride structure},\ }\href {https://doi.org/10.1016/0022-5088(76)90263-0} {\bibfield  {journal} {\bibinfo  {journal} {Journal of the Less Common Metals}\ }\textbf {\bibinfo {volume} {50}},\ \bibinfo {pages} {155} (\bibinfo {year} {1976})}\BibitemShut {NoStop}%
\bibitem [{\citenamefont {{Zimmerman}}(1972)}]{1972PhRvB...5.4704Z}%
  \BibitemOpen
  \bibfield  {author} {\bibinfo {author} {\bibfnamefont {W.~B.}\ \bibnamefont {{Zimmerman}}},\ }\bibfield  {title} {\bibinfo {title} {{Lattice-Constant Dependence on Isotopic Composition in the $\ce{^7Li}(\ce{H}, \ce{D})$ System}},\ }\href {https://doi.org/10.1103/PhysRevB.5.4704} {\bibfield  {journal} {\bibinfo  {journal} {\prb}\ }\textbf {\bibinfo {volume} {5}},\ \bibinfo {pages} {4704} (\bibinfo {year} {1972})}\BibitemShut {NoStop}%
\bibitem [{\citenamefont {Tsai}\ \emph {et~al.}(1956)\citenamefont {Tsai}, \citenamefont {Harris},\ and\ \citenamefont {Lassettre}}]{Tsai:1956}%
  \BibitemOpen
  \bibfield  {author} {\bibinfo {author} {\bibfnamefont {K.-R.}\ \bibnamefont {Tsai}}, \bibinfo {author} {\bibfnamefont {P.~M.}\ \bibnamefont {Harris}},\ and\ \bibinfo {author} {\bibfnamefont {E.~N.}\ \bibnamefont {Lassettre}},\ }\bibfield  {title} {\bibinfo {title} {{The Crystal Structure of Cesium Monoxide}},\ }\href {https://doi.org/10.1021/j150537a022} {\bibfield  {journal} {\bibinfo  {journal} {The Journal of Physical Chemistry}\ }\textbf {\bibinfo {volume} {60}},\ \bibinfo {pages} {338} (\bibinfo {year} {1956})}\BibitemShut {NoStop}%
\bibitem [{\citenamefont {Aleandri}\ and\ \citenamefont {McCarley}(1988)}]{Aleandri:1988}%
  \BibitemOpen
  \bibfield  {author} {\bibinfo {author} {\bibfnamefont {L.~E.}\ \bibnamefont {Aleandri}}\ and\ \bibinfo {author} {\bibfnamefont {R.~E.}\ \bibnamefont {McCarley}},\ }\bibfield  {title} {\bibinfo {title} {Hexagonal lithium molybdate, \ce{LiMoO2}: a close-packed layered structure with infinite molybdenum-molybdenum-bonded sheets},\ }\href {https://doi.org/10.1021/ic00279a021} {\bibfield  {journal} {\bibinfo  {journal} {Inorganic Chemistry}\ }\textbf {\bibinfo {volume} {27}},\ \bibinfo {pages} {1041} (\bibinfo {year} {1988})}\BibitemShut {NoStop}%
\bibitem [{\citenamefont {McCarroll}(1965)}]{McCarroll:1965}%
  \BibitemOpen
  \bibfield  {author} {\bibinfo {author} {\bibfnamefont {W.}~\bibnamefont {McCarroll}},\ }\bibfield  {title} {\bibinfo {title} {Chemical and structural characteristics of the potassium-cesium-antimony photocathode},\ }\href {https://doi.org/10.1016/0022-3697(65)90087-9} {\bibfield  {journal} {\bibinfo  {journal} {Journal of Physics and Chemistry of Solids}\ }\textbf {\bibinfo {volume} {26}},\ \bibinfo {pages} {191} (\bibinfo {year} {1965})}\BibitemShut {NoStop}%
\bibitem [{\citenamefont {Somer}\ \emph {et~al.}(2004)\citenamefont {Somer}, \citenamefont {Yarasik}, \citenamefont {Akselrud}, \citenamefont {Leoni}, \citenamefont {Rosner}, \citenamefont {Schnelle},\ and\ \citenamefont {Kniep}}]{Somer:2004}%
  \BibitemOpen
  \bibfield  {author} {\bibinfo {author} {\bibfnamefont {M.}~\bibnamefont {Somer}}, \bibinfo {author} {\bibfnamefont {A.}~\bibnamefont {Yarasik}}, \bibinfo {author} {\bibfnamefont {L.}~\bibnamefont {Akselrud}}, \bibinfo {author} {\bibfnamefont {S.}~\bibnamefont {Leoni}}, \bibinfo {author} {\bibfnamefont {H.}~\bibnamefont {Rosner}}, \bibinfo {author} {\bibfnamefont {W.}~\bibnamefont {Schnelle}},\ and\ \bibinfo {author} {\bibfnamefont {R.}~\bibnamefont {Kniep}},\ }\bibfield  {title} {\bibinfo {title} {{$Ae[\ce{Be2N2}]$: Nitridoberyllates of the Heavier Alkaline-Earth Metals}},\ }\href {https://doi.org/10.1002/anie.200352796} {\bibfield  {journal} {\bibinfo  {journal} {Angewandte Chemie International Edition}\ }\textbf {\bibinfo {volume} {43}},\ \bibinfo {pages} {1088} (\bibinfo {year} {2004})}\BibitemShut {NoStop}%
\bibitem [{\citenamefont {Ketblaar}\ and\ \citenamefont {van Walsem}(1938)}]{Ketblaar:1938}%
  \BibitemOpen
  \bibfield  {author} {\bibinfo {author} {\bibfnamefont {J.~A.~A.}\ \bibnamefont {Ketblaar}}\ and\ \bibinfo {author} {\bibfnamefont {J.~F.}\ \bibnamefont {van Walsem}},\ }\bibfield  {title} {\bibinfo {title} {{Die Krystallstruktur des Ammonium-, Kalium-, Rubidium- und C\"asiumpalladiumhexa-chlorids und -Bromids}},\ }\href {https://doi.org/10.1002/recl.19380570907} {\bibfield  {journal} {\bibinfo  {journal} {Recueil des Travaux Chimiques des Pays-Bas}\ }\textbf {\bibinfo {volume} {57}},\ \bibinfo {pages} {964} (\bibinfo {year} {1938})}\BibitemShut {NoStop}%
\bibitem [{\citenamefont {Broch}(1927)}]{Broch:1927}%
  \BibitemOpen
  \bibfield  {author} {\bibinfo {author} {\bibfnamefont {E.}~\bibnamefont {Broch}},\ }\bibfield  {title} {\bibinfo {title} {{Crystal Structure}},\ }\href@noop {} {\bibfield  {journal} {\bibinfo  {journal} {Zeitschrift fuer Physikalische Chemie}\ }\textbf {\bibinfo {volume} {127}},\ \bibinfo {pages} {446} (\bibinfo {year} {1927})}\BibitemShut {NoStop}%
\bibitem [{\citenamefont {{Lee}}\ \emph {et~al.}(2008)\citenamefont {{Lee}}, \citenamefont {{Bj{\"o}rling}}, \citenamefont {{Hauback}}, \citenamefont {{Utsumi}}, \citenamefont {{Moser}}, \citenamefont {{Bull}}, \citenamefont {{Nor{\'e}us}}, \citenamefont {{Sankey}},\ and\ \citenamefont {{H{\"a}ussermann}}}]{2008PhRvB..78s5209L}%
  \BibitemOpen
  \bibfield  {author} {\bibinfo {author} {\bibfnamefont {M.~H.}\ \bibnamefont {{Lee}}}, \bibinfo {author} {\bibfnamefont {T.}~\bibnamefont {{Bj{\"o}rling}}}, \bibinfo {author} {\bibfnamefont {B.~C.}\ \bibnamefont {{Hauback}}}, \bibinfo {author} {\bibfnamefont {T.}~\bibnamefont {{Utsumi}}}, \bibinfo {author} {\bibfnamefont {D.}~\bibnamefont {{Moser}}}, \bibinfo {author} {\bibfnamefont {D.}~\bibnamefont {{Bull}}}, \bibinfo {author} {\bibfnamefont {D.}~\bibnamefont {{Nor{\'e}us}}}, \bibinfo {author} {\bibfnamefont {O.~F.}\ \bibnamefont {{Sankey}}},\ and\ \bibinfo {author} {\bibfnamefont {U.}~\bibnamefont {{H{\"a}ussermann}}},\ }\bibfield  {title} {\bibinfo {title} {{Crystal structure, electronic structure, and vibrational properties of \ce{MAlSiH} ($\ce{M}=\ce{Ca},\ce{Sr},\ce{Ba}$): Hydrogenation-induced semiconductors from the \ce{AlB2}-type alloys \ce{MAlSi}}},\ }\href {https://doi.org/10.1103/PhysRevB.78.195209} {\bibfield  {journal} {\bibinfo  {journal} {\prb}\ }\textbf {\bibinfo {volume} {78}},\ \bibinfo
  {eid} {195209} (\bibinfo {year} {2008})}\BibitemShut {NoStop}%
\bibitem [{\citenamefont {Barker}\ \emph {et~al.}(2003{\natexlab{a}})\citenamefont {Barker}, \citenamefont {Saidi},\ and\ \citenamefont {Swoyer}}]{Barker:2003a}%
  \BibitemOpen
  \bibfield  {author} {\bibinfo {author} {\bibfnamefont {J.}~\bibnamefont {Barker}}, \bibinfo {author} {\bibfnamefont {M.~Y.}\ \bibnamefont {Saidi}},\ and\ \bibinfo {author} {\bibfnamefont {J.~L.}\ \bibnamefont {Swoyer}},\ }\bibfield  {title} {\bibinfo {title} {{Synthesis and Electrochemical Insertion Properties of the Layered \ce{Li_{x}MoO2} Phases ($x=0.74$, $0.85$, and $1.00$)}},\ }\href {https://doi.org/10.1149/1.1621288} {\bibfield  {journal} {\bibinfo  {journal} {Electrochemical and Solid-State Letters}\ }\textbf {\bibinfo {volume} {6}},\ \bibinfo {pages} {A252} (\bibinfo {year} {2003}{\natexlab{a}})}\BibitemShut {NoStop}%
\bibitem [{\citenamefont {Barker}\ \emph {et~al.}(2003{\natexlab{b}})\citenamefont {Barker}, \citenamefont {Saidi},\ and\ \citenamefont {Swoyer}}]{Barker:2003b}%
  \BibitemOpen
  \bibfield  {author} {\bibinfo {author} {\bibfnamefont {J.}~\bibnamefont {Barker}}, \bibinfo {author} {\bibfnamefont {M.}~\bibnamefont {Saidi}},\ and\ \bibinfo {author} {\bibfnamefont {J.}~\bibnamefont {Swoyer}},\ }\bibfield  {title} {\bibinfo {title} {{Lithium insertion properties of the layered \ce{LiMoO2} ($R\bar{3}m$) made by a novel carbothermal reduction method}},\ }\href {https://doi.org/https://doi.org/10.1016/S0167-2738(02)00875-5} {\bibfield  {journal} {\bibinfo  {journal} {Solid State Ionics}\ }\textbf {\bibinfo {volume} {158}},\ \bibinfo {pages} {261} (\bibinfo {year} {2003}{\natexlab{b}})}\BibitemShut {NoStop}%
\bibitem [{\citenamefont {Ben-Kamel}\ \emph {et~al.}(2012)\citenamefont {Ben-Kamel}, \citenamefont {Amdouni}, \citenamefont {Groult}, \citenamefont {Mauger}, \citenamefont {Zaghib},\ and\ \citenamefont {Julien}}]{ben-kamel:2012}%
  \BibitemOpen
  \bibfield  {author} {\bibinfo {author} {\bibfnamefont {K.}~\bibnamefont {Ben-Kamel}}, \bibinfo {author} {\bibfnamefont {N.}~\bibnamefont {Amdouni}}, \bibinfo {author} {\bibfnamefont {H.}~\bibnamefont {Groult}}, \bibinfo {author} {\bibfnamefont {A.}~\bibnamefont {Mauger}}, \bibinfo {author} {\bibfnamefont {K.}~\bibnamefont {Zaghib}},\ and\ \bibinfo {author} {\bibfnamefont {C.}~\bibnamefont {Julien}},\ }\bibfield  {title} {\bibinfo {title} {{Structural and electrochemical properties of \ce{LiMoO2}}},\ }\href {https://doi.org/https://doi.org/10.1016/j.jpowsour.2011.11.061} {\bibfield  {journal} {\bibinfo  {journal} {Journal of Power Sources}\ }\textbf {\bibinfo {volume} {202}},\ \bibinfo {pages} {314} (\bibinfo {year} {2012})}\BibitemShut {NoStop}%
\bibitem [{\citenamefont {{Flitcroft}}\ \emph {et~al.}(2024)\citenamefont {{Flitcroft}}, \citenamefont {{Althubiani}},\ and\ \citenamefont {{Skelton}}}]{2024JPEn....6b5011F}%
  \BibitemOpen
  \bibfield  {author} {\bibinfo {author} {\bibfnamefont {J.~M.}\ \bibnamefont {{Flitcroft}}}, \bibinfo {author} {\bibfnamefont {A.}~\bibnamefont {{Althubiani}}},\ and\ \bibinfo {author} {\bibfnamefont {J.~M.}\ \bibnamefont {{Skelton}}},\ }\bibfield  {title} {\bibinfo {title} {{Thermoelectric properties of the bismuth oxychalcogenides \ce{Bi2SO2}, \ce{Bi2SeO2} and \ce{Bi2TeO2}}},\ }\href {https://doi.org/10.1088/2515-7655/ad2afd} {\bibfield  {journal} {\bibinfo  {journal} {Journal of Physics: Energy}\ }\textbf {\bibinfo {volume} {6}},\ \bibinfo {eid} {025011} (\bibinfo {year} {2024})}\BibitemShut {NoStop}%
\bibitem [{\citenamefont {{Zhang}}\ \emph {et~al.}(2019)\citenamefont {{Zhang}}, \citenamefont {{Jiang}}, \citenamefont {{Song}}, \citenamefont {{Huang}}, \citenamefont {{He}}, \citenamefont {{Fang}}, \citenamefont {{Weng}},\ and\ \citenamefont {{Fang}}}]{2019Natur566475Z}%
  \BibitemOpen
  \bibfield  {author} {\bibinfo {author} {\bibfnamefont {T.}~\bibnamefont {{Zhang}}}, \bibinfo {author} {\bibfnamefont {Y.}~\bibnamefont {{Jiang}}}, \bibinfo {author} {\bibfnamefont {Z.}~\bibnamefont {{Song}}}, \bibinfo {author} {\bibfnamefont {H.}~\bibnamefont {{Huang}}}, \bibinfo {author} {\bibfnamefont {Y.}~\bibnamefont {{He}}}, \bibinfo {author} {\bibfnamefont {Z.}~\bibnamefont {{Fang}}}, \bibinfo {author} {\bibfnamefont {H.}~\bibnamefont {{Weng}}},\ and\ \bibinfo {author} {\bibfnamefont {C.}~\bibnamefont {{Fang}}},\ }\bibfield  {title} {\bibinfo {title} {{Catalogue of topological electronic materials}},\ }\href {https://doi.org/10.1038/s41586-019-0944-6} {\bibfield  {journal} {\bibinfo  {journal} {\nat}\ }\textbf {\bibinfo {volume} {566}},\ \bibinfo {pages} {475} (\bibinfo {year} {2019})},\ \Eprint {https://arxiv.org/abs/1807.08756} {arXiv:1807.08756 [cond-mat.mtrl-sci]} \BibitemShut {NoStop}%
\bibitem [{\citenamefont {Hochberg}\ \emph {et~al.}()\citenamefont {Hochberg}, \citenamefont {Lehmann} \emph {et~al.}}]{futureDFT}%
  \BibitemOpen
  \bibfield  {author} {\bibinfo {author} {\bibfnamefont {Y.}~\bibnamefont {Hochberg}}, \bibinfo {author} {\bibfnamefont {B.~V.}\ \bibnamefont {Lehmann}}, \emph {et~al.},\ }\href@noop {} {\bibinfo  {journal} {to appear}\ }\BibitemShut {NoStop}%
\bibitem [{\citenamefont {Cook}\ \emph {et~al.}(2024)\citenamefont {Cook}, \citenamefont {Blanco},\ and\ \citenamefont {Smirnov}}]{Cook:2024cgm}%
  \BibitemOpen
\bibfield  {journal} {  }\bibfield  {author} {\bibinfo {author} {\bibfnamefont {C.}~\bibnamefont {Cook}}, \bibinfo {author} {\bibfnamefont {C.}~\bibnamefont {Blanco}},\ and\ \bibinfo {author} {\bibfnamefont {J.}~\bibnamefont {Smirnov}},\ }\bibfield  {title} {\bibinfo {title} {{Deep learning optimal molecular scintillators for dark matter direct detection}},\ }\href@noop {} {\  (\bibinfo {year} {2024})},\ \Eprint {https://arxiv.org/abs/2501.00091} {arXiv:2501.00091 [hep-ph]} \BibitemShut {NoStop}%
\bibitem [{\citenamefont {{Benjamin V. Lehmann}}(2025)}]{montecolor}%
  \BibitemOpen
  \bibfield  {author} {\bibinfo {author} {\bibnamefont {{Benjamin V. Lehmann}}},\ }\href {https://github.com/benvlehmann/montecolor} {\bibinfo {title} {{MonteColor}}} (\bibinfo {year} {2025})\BibitemShut {NoStop}%
\bibitem [{\citenamefont {Luo}\ \emph {et~al.}(2006)\citenamefont {Luo}, \citenamefont {Cui},\ and\ \citenamefont {Li}}]{Luo:2006}%
  \BibitemOpen
  \bibfield  {author} {\bibinfo {author} {\bibfnamefont {M.~R.}\ \bibnamefont {Luo}}, \bibinfo {author} {\bibfnamefont {G.}~\bibnamefont {Cui}},\ and\ \bibinfo {author} {\bibfnamefont {C.}~\bibnamefont {Li}},\ }\bibfield  {title} {\bibinfo {title} {{Uniform colour spaces based on CIECAM02 colour appearance model}},\ }\href {https://doi.org/10.1002/col.20227} {\bibfield  {journal} {\bibinfo  {journal} {Color Research \& Application}\ }\textbf {\bibinfo {volume} {31}},\ \bibinfo {pages} {320} (\bibinfo {year} {2006})}\BibitemShut {NoStop}%
\bibitem [{\citenamefont {Perdew}\ \emph {et~al.}(1996)\citenamefont {Perdew}, \citenamefont {Burke},\ and\ \citenamefont {Ernzerhof}}]{Perdew:1996pki}%
  \BibitemOpen
  \bibfield  {author} {\bibinfo {author} {\bibfnamefont {J.~P.}\ \bibnamefont {Perdew}}, \bibinfo {author} {\bibfnamefont {K.}~\bibnamefont {Burke}},\ and\ \bibinfo {author} {\bibfnamefont {M.}~\bibnamefont {Ernzerhof}},\ }\bibfield  {title} {\bibinfo {title} {{Generalized Gradient Approximation Made Simple}},\ }\href {https://doi.org/10.1103/PhysRevLett.77.3865} {\bibfield  {journal} {\bibinfo  {journal} {Phys. Rev. Lett.}\ }\textbf {\bibinfo {volume} {77}},\ \bibinfo {pages} {3865} (\bibinfo {year} {1996})}\BibitemShut {NoStop}%
\bibitem [{\citenamefont {{Kingsbury}}\ \emph {et~al.}(2022)\citenamefont {{Kingsbury}}, \citenamefont {{Gupta}}, \citenamefont {{Bartel}}, \citenamefont {{Munro}}, \citenamefont {{Dwaraknath}}, \citenamefont {{Horton}},\ and\ \citenamefont {{Persson}}}]{2022PhRvM...6a3801K}%
  \BibitemOpen
  \bibfield  {author} {\bibinfo {author} {\bibfnamefont {R.}~\bibnamefont {{Kingsbury}}}, \bibinfo {author} {\bibfnamefont {A.~S.}\ \bibnamefont {{Gupta}}}, \bibinfo {author} {\bibfnamefont {C.~J.}\ \bibnamefont {{Bartel}}}, \bibinfo {author} {\bibfnamefont {J.~M.}\ \bibnamefont {{Munro}}}, \bibinfo {author} {\bibfnamefont {S.}~\bibnamefont {{Dwaraknath}}}, \bibinfo {author} {\bibfnamefont {M.}~\bibnamefont {{Horton}}},\ and\ \bibinfo {author} {\bibfnamefont {K.~A.}\ \bibnamefont {{Persson}}},\ }\bibfield  {title} {\bibinfo {title} {{Performance comparison of r$^{2}$SCAN and SCAN metaGGA density functionals for solid materials via an automated, high-throughput computational workflow}},\ }\href {https://doi.org/10.1103/PhysRevMaterials.6.013801} {\bibfield  {journal} {\bibinfo  {journal} {Physical Review Materials}\ }\textbf {\bibinfo {volume} {6}},\ \bibinfo {eid} {013801} (\bibinfo {year} {2022})}\BibitemShut {NoStop}%
\bibitem [{\citenamefont {Furness}\ \emph {et~al.}(2020)\citenamefont {Furness}, \citenamefont {Kaplan}, \citenamefont {Ning}, \citenamefont {Perdew},\ and\ \citenamefont {Sun}}]{Furness:2020}%
  \BibitemOpen
  \bibfield  {author} {\bibinfo {author} {\bibfnamefont {J.~W.}\ \bibnamefont {Furness}}, \bibinfo {author} {\bibfnamefont {A.~D.}\ \bibnamefont {Kaplan}}, \bibinfo {author} {\bibfnamefont {J.}~\bibnamefont {Ning}}, \bibinfo {author} {\bibfnamefont {J.~P.}\ \bibnamefont {Perdew}},\ and\ \bibinfo {author} {\bibfnamefont {J.}~\bibnamefont {Sun}},\ }\bibfield  {title} {\bibinfo {title} {{Accurate and Numerically Efficient r$^2$SCAN Meta-Generalized Gradient Approximation}},\ }\href {https://doi.org/10.1021/acs.jpclett.0c02405} {\bibfield  {journal} {\bibinfo  {journal} {The Journal of Physical Chemistry Letters}\ }\textbf {\bibinfo {volume} {11}},\ \bibinfo {pages} {8208} (\bibinfo {year} {2020})}\BibitemShut {NoStop}%
\bibitem [{\citenamefont {Heyd}\ \emph {et~al.}(2003)\citenamefont {Heyd}, \citenamefont {Scuseria},\ and\ \citenamefont {Ernzerhof}}]{Heyd:2003rlw}%
  \BibitemOpen
  \bibfield  {author} {\bibinfo {author} {\bibfnamefont {J.}~\bibnamefont {Heyd}}, \bibinfo {author} {\bibfnamefont {G.~E.}\ \bibnamefont {Scuseria}},\ and\ \bibinfo {author} {\bibfnamefont {M.}~\bibnamefont {Ernzerhof}},\ }\bibfield  {title} {\bibinfo {title} {{Hybrid functionals based on a screened Coulomb potential}},\ }\href {https://doi.org/10.1063/1.1564060} {\bibfield  {journal} {\bibinfo  {journal} {J. Chem. Phys.}\ }\textbf {\bibinfo {volume} {118}},\ \bibinfo {pages} {8207} (\bibinfo {year} {2003})}\BibitemShut {NoStop}%
\bibitem [{\citenamefont {{Krukau}}\ \emph {et~al.}(2006)\citenamefont {{Krukau}}, \citenamefont {{Vydrov}}, \citenamefont {{Izmaylov}},\ and\ \citenamefont {{Scuseria}}}]{2006JChPh.125v4106K}%
  \BibitemOpen
  \bibfield  {author} {\bibinfo {author} {\bibfnamefont {A.~V.}\ \bibnamefont {{Krukau}}}, \bibinfo {author} {\bibfnamefont {O.~A.}\ \bibnamefont {{Vydrov}}}, \bibinfo {author} {\bibfnamefont {A.~F.}\ \bibnamefont {{Izmaylov}}},\ and\ \bibinfo {author} {\bibfnamefont {G.~E.}\ \bibnamefont {{Scuseria}}},\ }\bibfield  {title} {\bibinfo {title} {{Influence of the exchange screening parameter on the performance of screened hybrid functionals}},\ }\href {https://doi.org/10.1063/1.2404663} {\bibfield  {journal} {\bibinfo  {journal} {\jcp}\ }\textbf {\bibinfo {volume} {125}},\ \bibinfo {pages} {224106} (\bibinfo {year} {2006})}\BibitemShut {NoStop}%
\bibitem [{\citenamefont {{Zhao}}\ \emph {et~al.}()\citenamefont {{Zhao}}, \citenamefont {{Yang}}, \citenamefont {{Kaplan}},\ and\ \citenamefont {{Persson}}}]{Zhao:2025}%
  \BibitemOpen
  \bibfield  {author} {\bibinfo {author} {\bibfnamefont {W.}~\bibnamefont {{Zhao}}}, \bibinfo {author} {\bibfnamefont {R.~X.}\ \bibnamefont {{Yang}}}, \bibinfo {author} {\bibfnamefont {A.~D.}\ \bibnamefont {{Kaplan}}},\ and\ \bibinfo {author} {\bibfnamefont {K.~A.}\ \bibnamefont {{Persson}}},\ }\bibfield  {title} {\bibinfo {title} {{Accelerated discovery of cost-effective photoabsorber materials for near-infrared ($\lambda=\qty{1600}{\nano\meter}$) photodetector applications}},\ }\href@noop {} {\ }\Eprint {https://arxiv.org/abs/2504.16317} {arXiv:2504.16317 [cond-mat.mtrl-sci]} \BibitemShut {NoStop}%
\bibitem [{\citenamefont {Carrazza}\ and\ \citenamefont {Cruz-Martinez}(2020)}]{Carrazza:2020rdn}%
  \BibitemOpen
  \bibfield  {author} {\bibinfo {author} {\bibfnamefont {S.}~\bibnamefont {Carrazza}}\ and\ \bibinfo {author} {\bibfnamefont {J.~M.}\ \bibnamefont {Cruz-Martinez}},\ }\bibfield  {title} {\bibinfo {title} {{VegasFlow: accelerating Monte Carlo simulation across multiple hardware platforms}},\ }\href {https://doi.org/10.1016/j.cpc.2020.107376} {\bibfield  {journal} {\bibinfo  {journal} {Comput. Phys. Commun.}\ }\textbf {\bibinfo {volume} {254}},\ \bibinfo {pages} {107376} (\bibinfo {year} {2020})},\ \Eprint {https://arxiv.org/abs/2002.12921} {arXiv:2002.12921 [physics.comp-ph]} \BibitemShut {NoStop}%
\bibitem [{\citenamefont {Cruz-Martinez}\ and\ \citenamefont {Carrazza}(2020)}]{vegasflow_package}%
  \BibitemOpen
  \bibfield  {author} {\bibinfo {author} {\bibfnamefont {J.}~\bibnamefont {Cruz-Martinez}}\ and\ \bibinfo {author} {\bibfnamefont {S.}~\bibnamefont {Carrazza}},\ }\href {https://doi.org/10.5281/zenodo.3691926} {\bibinfo {title} {N3pdf/vegasflow: vegasflow v1.0}} (\bibinfo {year} {2020})\BibitemShut {NoStop}%
\bibitem [{\citenamefont {Barak}\ \emph {et~al.}(2020)\citenamefont {Barak} \emph {et~al.}}]{Barak:2020fql}%
  \BibitemOpen
  \bibfield  {author} {\bibinfo {author} {\bibfnamefont {L.}~\bibnamefont {Barak}} \emph {et~al.} (\bibinfo {collaboration} {SENSEI}),\ }\bibfield  {title} {\bibinfo {title} {{SENSEI: Direct-Detection Results on sub-GeV Dark Matter from a New Skipper-CCD}},\ }\href {https://doi.org/10.1103/PhysRevLett.125.171802} {\bibfield  {journal} {\bibinfo  {journal} {Phys. Rev. Lett.}\ }\textbf {\bibinfo {volume} {125}},\ \bibinfo {pages} {171802} (\bibinfo {year} {2020})},\ \Eprint {https://arxiv.org/abs/2004.11378} {arXiv:2004.11378 [astro-ph.CO]} \BibitemShut {NoStop}%
\bibitem [{\citenamefont {Amaral}\ \emph {et~al.}(2020)\citenamefont {Amaral} \emph {et~al.}}]{Amaral:2020ryn}%
  \BibitemOpen
  \bibfield  {author} {\bibinfo {author} {\bibfnamefont {D.}~\bibnamefont {Amaral}} \emph {et~al.} (\bibinfo {collaboration} {SuperCDMS}),\ }\bibfield  {title} {\bibinfo {title} {{Constraints on low-mass, relic dark matter candidates from a surface-operated SuperCDMS single-charge sensitive detector}},\ }\href {https://doi.org/10.1103/PhysRevD.102.091101} {\bibfield  {journal} {\bibinfo  {journal} {Phys. Rev. D}\ }\textbf {\bibinfo {volume} {102}},\ \bibinfo {pages} {091101} (\bibinfo {year} {2020})},\ \Eprint {https://arxiv.org/abs/2005.14067} {arXiv:2005.14067 [hep-ex]} \BibitemShut {NoStop}%
\bibitem [{\citenamefont {Aguilar-Arevalo}\ \emph {et~al.}(2019)\citenamefont {Aguilar-Arevalo} \emph {et~al.}}]{Aguilar-Arevalo:2019wdi}%
  \BibitemOpen
  \bibfield  {author} {\bibinfo {author} {\bibfnamefont {A.}~\bibnamefont {Aguilar-Arevalo}} \emph {et~al.} (\bibinfo {collaboration} {DAMIC}),\ }\bibfield  {title} {\bibinfo {title} {{Constraints on Light Dark Matter Particles Interacting with Electrons from DAMIC at SNOLAB}},\ }\href {https://doi.org/10.1103/PhysRevLett.123.181802} {\bibfield  {journal} {\bibinfo  {journal} {Phys. Rev. Lett.}\ }\textbf {\bibinfo {volume} {123}},\ \bibinfo {pages} {181802} (\bibinfo {year} {2019})},\ \Eprint {https://arxiv.org/abs/1907.12628} {arXiv:1907.12628 [astro-ph.CO]} \BibitemShut {NoStop}%
\bibitem [{\citenamefont {Essig}\ \emph {et~al.}(2017)\citenamefont {Essig}, \citenamefont {Volansky},\ and\ \citenamefont {Yu}}]{Essig:2017kqs}%
  \BibitemOpen
  \bibfield  {author} {\bibinfo {author} {\bibfnamefont {R.}~\bibnamefont {Essig}}, \bibinfo {author} {\bibfnamefont {T.}~\bibnamefont {Volansky}},\ and\ \bibinfo {author} {\bibfnamefont {T.-T.}\ \bibnamefont {Yu}},\ }\bibfield  {title} {\bibinfo {title} {{New Constraints and Prospects for sub-GeV Dark Matter Scattering off Electrons in Xenon}},\ }\href {https://doi.org/10.1103/PhysRevD.96.043017} {\bibfield  {journal} {\bibinfo  {journal} {Phys. Rev. D}\ }\textbf {\bibinfo {volume} {96}},\ \bibinfo {pages} {043017} (\bibinfo {year} {2017})},\ \Eprint {https://arxiv.org/abs/1703.00910} {arXiv:1703.00910 [hep-ph]} \BibitemShut {NoStop}%
\bibitem [{\citenamefont {Agnes}\ \emph {et~al.}(2018)\citenamefont {Agnes} \emph {et~al.}}]{Agnes:2018oej}%
  \BibitemOpen
  \bibfield  {author} {\bibinfo {author} {\bibfnamefont {P.}~\bibnamefont {Agnes}} \emph {et~al.} (\bibinfo {collaboration} {DarkSide}),\ }\bibfield  {title} {\bibinfo {title} {{Constraints on Sub-GeV Dark-Matter\textendash{}Electron Scattering from the DarkSide-50 Experiment}},\ }\href {https://doi.org/10.1103/PhysRevLett.121.111303} {\bibfield  {journal} {\bibinfo  {journal} {Phys. Rev. Lett.}\ }\textbf {\bibinfo {volume} {121}},\ \bibinfo {pages} {111303} (\bibinfo {year} {2018})},\ \Eprint {https://arxiv.org/abs/1802.06998} {arXiv:1802.06998 [astro-ph.CO]} \BibitemShut {NoStop}%
\bibitem [{\citenamefont {Aprile}\ \emph {et~al.}(2019)\citenamefont {Aprile} \emph {et~al.}}]{Aprile:2019xxb}%
  \BibitemOpen
  \bibfield  {author} {\bibinfo {author} {\bibfnamefont {E.}~\bibnamefont {Aprile}} \emph {et~al.} (\bibinfo {collaboration} {XENON}),\ }\bibfield  {title} {\bibinfo {title} {{Light Dark Matter Search with Ionization Signals in XENON1T}},\ }\href {https://doi.org/10.1103/PhysRevLett.123.251801} {\bibfield  {journal} {\bibinfo  {journal} {Phys. Rev. Lett.}\ }\textbf {\bibinfo {volume} {123}},\ \bibinfo {pages} {251801} (\bibinfo {year} {2019})},\ \Eprint {https://arxiv.org/abs/1907.11485} {arXiv:1907.11485 [hep-ex]} \BibitemShut {NoStop}%
\end{thebibliography}%

\end{document}